\documentclass[prl,aps,twocolumn,groupedaddress,floats,final,superscriptaddress,nopacks]{revtex4-2}
\usepackage{graphicx}
\usepackage[utf8x]{inputenc}
\usepackage{lineno}
\usepackage{color}
\usepackage{subfigure}
\usepackage{latexsym}
\usepackage{epsfig}
\usepackage{psfrag}
\usepackage{amsmath,amsfonts,amssymb,bm,ulem}
\usepackage{float}
\usepackage{ulem}

\begin{document}

\title{Dynamic Response of an Electron Gas: Towards the Exact Exchange-Correlation Kernel}

\author{James P. F. LeBlanc}
\affiliation{Department of Physics and Physical Oceanography, Memorial University of Newfoundland,
St. John's, Newfoundland \& Labrador, Canada A1B 3X7}
\author{Kun Chen}
\affiliation{Center for Computational Quantum Physics, Flatiron Institute, 162 Fifth Avenue, New York, NY 1001, USA}
\author{Kristjan Haule}
\affiliation{Department of Physics and Astronomy, Rutgers University, Piscataway, NJ 08854, USA}
\author{Nikolay V. Prokof'ev}
\affiliation{Department of Physics, University of Massachusetts, Amherst, MA 01003, USA}
\author{Igor S. Tupitsyn}
\email{itupitsyn@physics.umass.edu}
\affiliation{Department of Physics, University of Massachusetts, Amherst, MA 01003, USA}

\begin{abstract}
Precise calculations of dynamics in the homogeneous electron gas (jellium model) are of fundamental importance for design and characterization of new materials. We introduce a diagrammatic Monte Carlo technique based on algorithmic Matsubara integration that allows us to compute frequency and momentum resolved finite temperature response directly in the real frequency domain using series of connected Feynman diagrams. The data for charge response at moderate electron density are used to extract the frequency dependence of the exchange-correlation kernel at finite momenta and temperature. These results are as important for development of the time-dependent density functional theory for materials dynamics as ground state energies are for the density functional theory.
\end{abstract}

\maketitle

%%%%%%%%%%%%%%%%%%%%%%%%%%%%%%%%%%%%%%%%%%%%%%%%%%%%%%%%%%%%%%%%%%%%%%%%%%%%%%%%%%

\textbf{Introduction.} To predict functional behavior of new materials the knowledge of their dynamic response functions at finite temperature is crucial. In this context, the jellium model plays a special role both as a paradigmatic system for understanding the physics of the electron liquid in solids \cite{Ceperley80,Utsumi,Moroni,Bowen,Perdew92,Chachiyo} as well as being the key element in the formulation of the time-dependent density functional theory (TDDFT) \cite{Runge1984,Kohn1985}. The model itself is defined by the interacting homogeneous electron gas stabilized by the positive neutralizing background.

Typically, finite-$T$ many-body calculations are performed in the Matsubara representation, i.e. on the imaginary time or frequency axis \cite{Matsubara1955}, and real-frequency results are recovered only by performing a numerical analytic continuation (NAC). This poses a major problem for accurate theoretical descriptions of experimentally relevant observables because even when the imaginary axis data are known with very high accuracy, the NAC will not resolve the fine spectral features at finite frequency or correctly reproduce complex spectra unless enough known constrains are imposed in the analysis, which is seldom possible \cite{Goulko2017}. Until recently, this infamous problem was considered unavoidable.

Recent breakthroughs in solving the jellium model by the diagrammatic Monte Carlo (DiagMC) method
in the Matsubara representation \cite{Chen2019,Haule2022} and applying the Algorithmic Matsubara integration (AMI) to the Hubbard model \cite{LeBlanc2019,LeBlanc2020a,LeBlanc2020b}
(see also Refs.~\cite{FerreroRT,Ferrero2020}) paved the road for accurate studies of finite-$T$ dynamic response in jellium. In this work, we combine these two breakthroughs by developing the Algorithmic Matsubara-diagrammatic Monte Carlo (ADiagMC) technique to study dynamic properties of jellium without employing the NAC. In particular, we demonstrate that finite-T computations of the momentum and real frequency resolved dielectric functions and exchange-correlation kernels are now possible.

In the DiagMC approach for jellium \cite{Chen2019,Haule2022} all listed connected diagrams of a given order $N$ are grouped together with the properly optimized internal integration variables to suppress variance from sign-canceling contributions. The ADiagMC technique lists all diagrams of order $N$, performs the analytic summation over internal Matsubara frequencies for every listed diagram \cite{LeBlanc2019,LeBlanc2020a}, and stochastically samples integrals over internal momenta. The DiagMC approach works directly in the thermodynamic limit \cite{DiagMC-1,DiagMC-2}, does not suffer from the conventional notorious fermionic sign problem \cite{Signproblem}, and can be applied to systems with arbitrary dispersion relations and shapes of the interaction potential \cite{DiagMC-1,DiagMC-2,SimonsHydr,Dirac2017,Mishch2021}. The ADiagMC technique works in the same way.

%%%%%%%%%%%%%%%%%%%%%%%%%%%%%%%%%%%%%%%%%%%%%%%%%%%%%%%%%%%%%%%%%%%%%%%%%%%%%%%%%%

\smallskip

\noindent \textbf{Real frequency technique for jellium.} The Hamiltonian of the jellium model is defined by
\begin{equation}
H=\sum_i \frac{k_i^2}{2m} + \sum_{i<j} \frac{e^2}{|\mathbf{r}_i -\mathbf{r}_j|} - \mu N ,
\label{jellium}
\end{equation}
with $m$ the electron mass, $\mu$ the chemical potential, and $e$ the electron charge.
We use the inverse Fermi momentum, $1/k_F$, and Fermi energy, $\varepsilon_F = k_F^2/2m$,
as units of length and energy, respectively; the definition of the Coulomb parameter $r_s$
in terms of the particle number density, $\rho = k_F^3/3\pi^2$, and Bohr radius, $a_B=1/me^2$,
is standard: $4\pi r_s^3/3 = 1/ \rho a_B^3$.

The starting point for all considerations is connected Feynman diagrams for a system of interacting fermions written in the momentum-frequency representation. To account for correlations, an expansion is performed in terms of renormalized single particle propagators and screened effective interactions~\cite{Chen2019,ShiftAct} and contributions from all diagrams up to some maximum expansion order $N$ are computed. In this work we focus on the charge response and compute the polarization function $\Pi({\bf Q},\Omega,T)$. The dielectric function, $\epsilon({\bf Q},\Omega,T)$, is related to $\Pi$ in the standard way, $\epsilon = 1 - V_0 \Pi$, where $V_0=4\pi e^2/Q^2$ is the bare Coulomb interaction and $\Omega$ is real frequency. Instead of computing $\Pi$ in the Matsubara representation and then applying NAC, the AMI technique \cite{LeBlanc2019,LeBlanc2020a} symbolically generates \textit{analytic} expressions for sums over all internal Matsubara frequencies ($\omega_s=2 \pi T (s+1/2)$ for fermions and $2 \pi T s$ for bosons) and performs the ``Wick rotation" of external frequency from imaginary to real frequency axis analytically by simply substituting $i\Omega_s$ with $\Omega + i \eta$ (for more details see Note II in the Supplemental material \cite{SM}). This protocol works at finite temperature and eliminates all problems associated with NAC. It utilizes expressions that are analytic functions of temperature and thus any temperature is potentially accessible.

The ability to account for the high-order Feynman diagrams is important for estimating the accuracy of calculation, and in the Coulomb system this is only possible by incorporating screening into a new non-interacting action $S_0$ using shifted and homotopic action tools \cite{ShiftAct,homotopy}. The idea is to rewrite the system's action identically as $S=S_0+\Delta S$, with $\Delta S$ composed of interactions and the so-called ``counter-terms'' that partially or completely compensate contributions from diagrams generated by interactions. Our choice is to replace the Coulomb interaction $V_0$ with the Yukawa one, $Y=4 \pi e^2/(q^2 + \kappa^2)$, where $\kappa$ is some screening momentum. To understand how compensation for an arbitrary $\kappa$ works, consider an effective Coulomb potential, $U = V_0/[1-V_0\Pi]$, and rewrite it identically as $U \equiv Y/[1-Y(\Pi + \kappa^2/4 \pi e^2)]$. Thus, if the bare Coulomb potential $V_0(q)$ is replaced with $Y(q)$, then the diagrammatic expansion in powers of $Y$ should be augmented with the ``polarization'' counter-term $\kappa^2/4 \pi e^2$. The value of $\kappa$ can be optimized order-by-order for faster convergence \cite{Chen2019}. For dynamic properties one should also pay attention to causality \cite{TTKP2021}. In this work we chose $\kappa$=1 from a broad extremum of the static charge polarization, $\Pi(q=0,\omega=0,\kappa)$, where it remains nearly constant (for more details see Ref.~\cite{Chen2019}).

In addition, to ensure that the expansion is performed at constant electron density $\rho$ (fixed by the value of the Coulomb parameter $r_s$) we employ the ``chemical potential'' counter-terms. Even if the chemical potential is fine-tuned to reproduce $\rho$ at the self-consistent Hartree-Fock level, higher order self-energy corrections would still result in the density changes. The standard renormalization scheme is to introduce counter-terms based on the chemical potential shifts $\delta \mu_n$ for proper self-energy diagrams of the order $n$ such that the series for the Green's function satisfy $2 \sum_k  n_{\mathbf k} = \rho$ at each order of expansion.

We expand on top of the self-consistent Hartree-Fock (HF) solution for the Green's function: $G^{-1} = G^{-1}_0 - \Sigma_F[G]$, where $G_0$ is the bare Green's function. This solution is based on the Fock diagram for the proper self-energy (Hartree diagram is canceled by charge neutrality): $\Sigma_F(\mathbf{k}) = \sum_{\mathbf q} Y(\mathbf{q}) n(\epsilon_{\mathbf{k} - \mathbf{q}})$, where $n(\epsilon_{\mathbf{k}}) \equiv G(\mathbf{k}, \tau=-0)$ are finite-temperature Fermi occupation numbers. Note that $G = (\omega - k^2/2m - \Sigma_F(\mathbf{k}) + \mu)^{-1} \equiv (\omega - \epsilon_{\mathbf{k}})^{-1}$ has the same simple pole structure as $G_0$. By incorporating all Fock diagrams into $G$ we simplify the series expansion by omitting all diagrams with Fock type self-energy insertions.

Each diagram for the polarization function $\Pi$ is characterized by three integers $a,b,c$ defining the order of expansion $N = a+b+\sum_{k=1}^{c}r_k $: $a$ is the number of independent internal momenta, $b$ is the number of polarization counter-terms, and $c$ is the number of self-energy counter-terms (the minimal value of $r$ for self-energy counter-term is $r=2$ because Fock diagrams are excluded, for more details see Note I in the Supplemental material \cite{SM}). In what follows, the ``N-th order result'' means that all diagrams up to the N-th order are included in the answer.

In the rest of the paper we demonstrate how our technique works for the jellium model and allows us to produce unique results for dynamic response at finite temperature. All results in the main text are based on the $N=3$ calculations for the polarization function $\Pi$ with selected $N=4$ and $N=5$ calculations used to estimate the accuracy bounds, see Fig.~\ref{Fig5} below (and, also, Note IV in the Supplemental material \cite{SM}).

%%%%%%%%%%%%%%%%%%%%%%%%%%%%%%%%%%%%%%%%%%%%%%%%%%%%%%%%%%%%%%%%%%%%%%%%%%%%%%%%
\smallskip

\begin{figure}[t]
%\vspace{-2mm}
\subfigure{\includegraphics[scale=0.25]{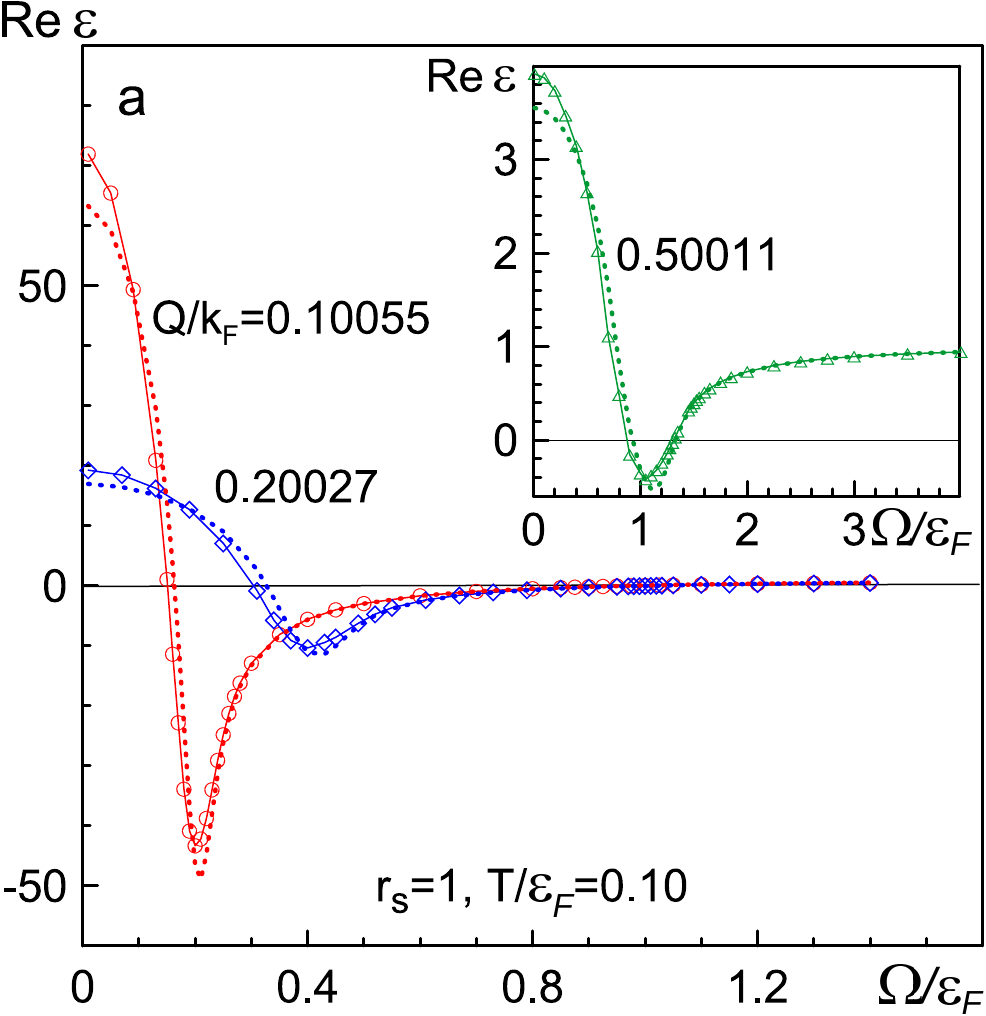}}
\subfigure{\includegraphics[scale=0.25]{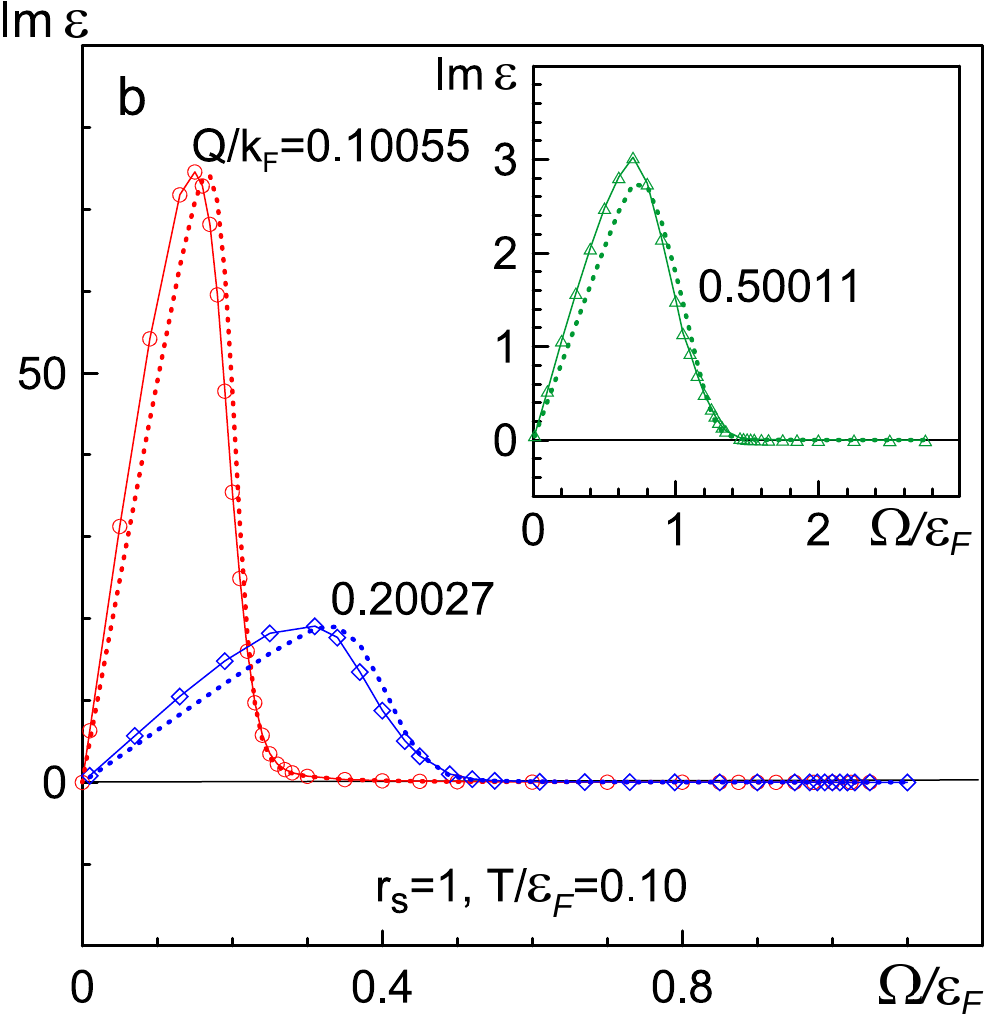}}
\subfigure{\includegraphics[scale=0.25]{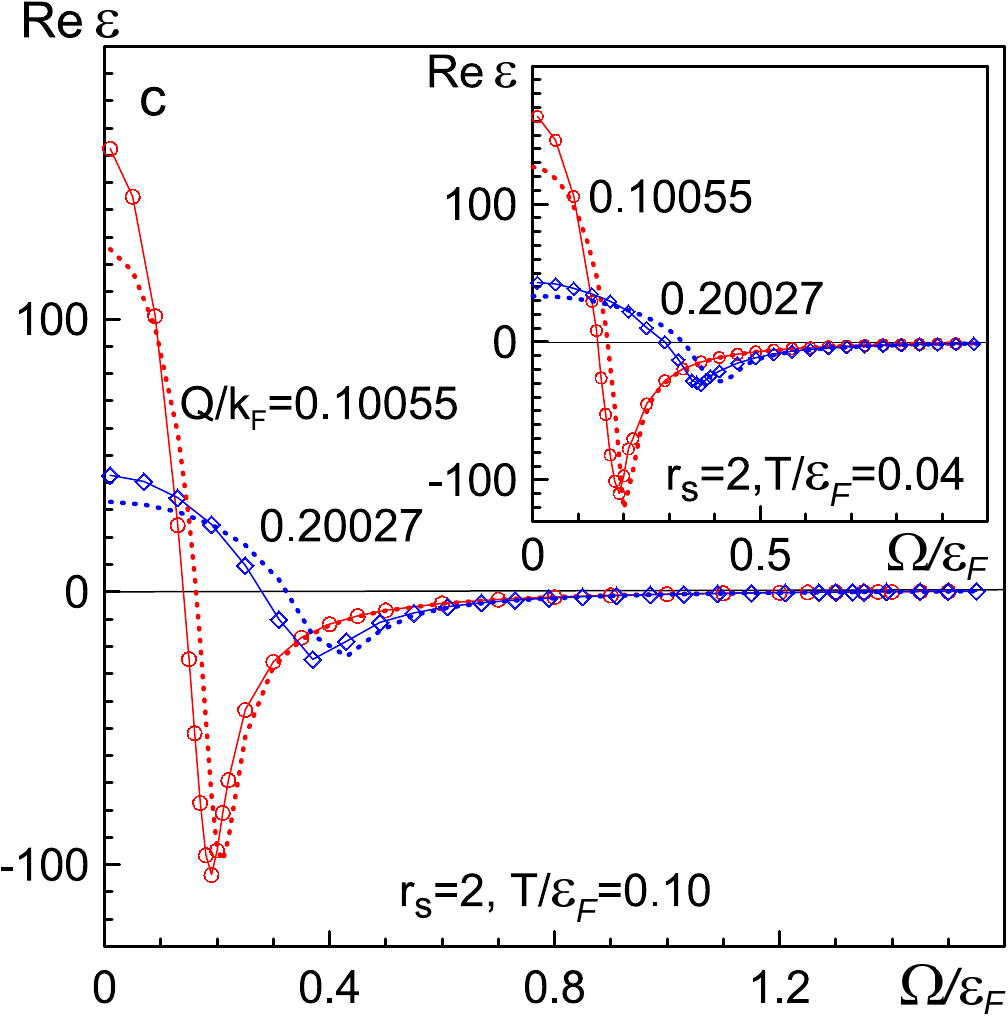}}
\subfigure{\includegraphics[scale=0.25]{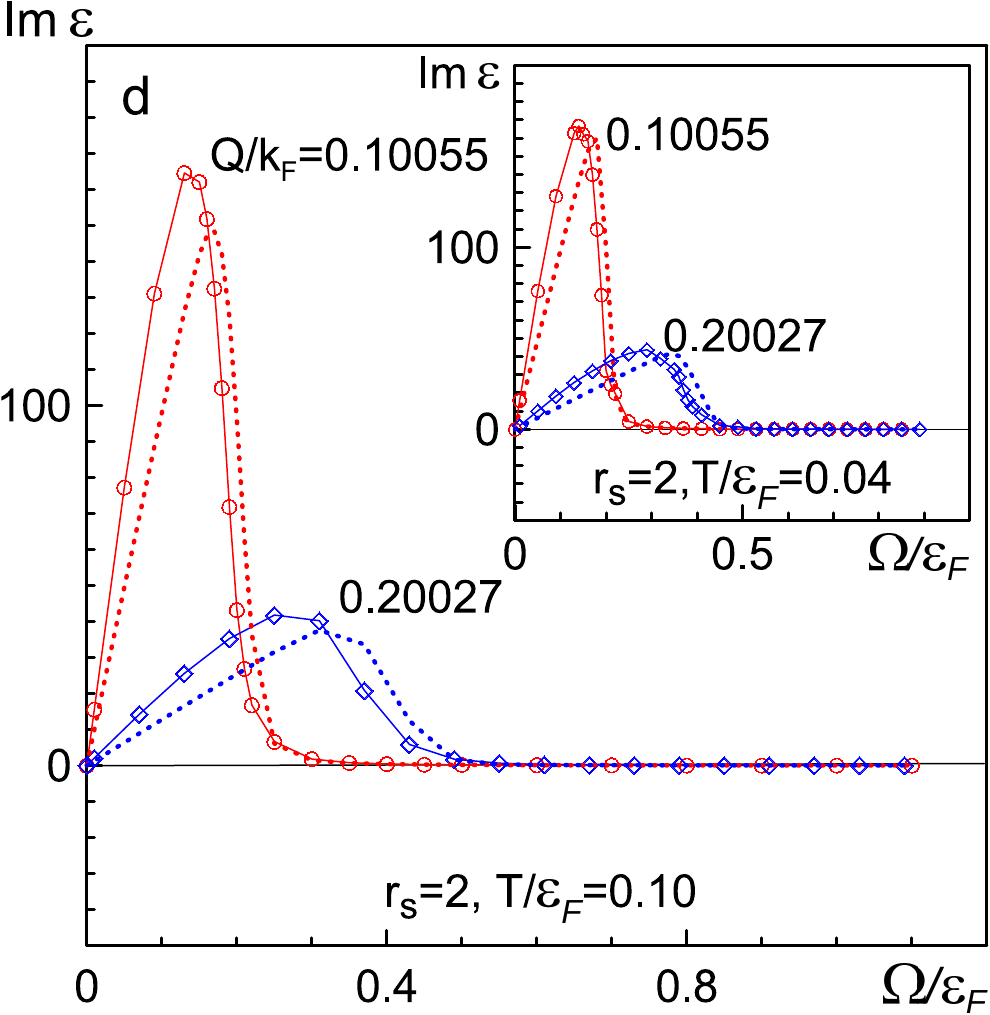}}
%\vspace{-2mm}
\caption{Real and imaginary parts of the dielectric function as functions of frequency at different
momenta and temperatures for $r_s = 1$ (a,b) and $r_s=2$ (c,d). Solid curves with symbols: ADiagMC results. Dotted curves: RPA results for the same parameter sets including $T$ and $\eta$ values.
Insets in (a) and (b) present results for larger momentum transfer $Q$. Insets in (c) and (d) show the effect of lowering the temperature $T$. Errors are within the symbol sizes.}
\label{Fig1}
\end{figure}

\noindent
\textbf{Dielectric function.} In Fig.~\ref{Fig1} we compare our results for the dielectric function
with the leading-order random phase approximation (RPA) for the same set of parameters. The $T$ and $\eta$ dependent polarization function within RPA is given by
\begin{equation}
\Pi_{RPA}(\textbf{Q},\Omega,T) = - 2 \int \frac{d^3p}{(2\pi )^{3}}
\frac{ n(\varepsilon_{{\bf p} + {\bf Q}})-n(\varepsilon_{{\bf p}} ) }
     {\Omega - \varepsilon_{{\bf p} + {\bf Q}} + \varepsilon_{{\bf p}} + i \eta},
\label{RPA}
\end{equation}
where $\varepsilon_{{\bf p}} = p^2/2m - \mu$ is the bare electron dispersion. As expected, corrections to RPA grow with the value of $r_s$, and can exceed $20 \%$ for some points at $r_s=2$. Zeros of $\mathrm{Re} \; \epsilon$ at frequencies $\Omega > v_F Q$, where $v_F$ is the Fermi velocity (in our units $v_F = 2$), reveal the collective plasmon mode with dispersion $\omega_{\mathrm{pl}}({\bf Q},T)$. At momentum $Q_m$ the plasmon branch and the electron-hole ($e-h$) continuum merge; the inset in Fig.~\ref{Fig1}a shows two close zeros of $\mathrm{Re} \; \epsilon$ for $Q$ slightly below $Q_m$. The value of $Q_m$ increases with $r_s$ and can be approximately determined from the condition $\omega_{\mathrm{pl}} (Q_m) = \xi(k_F+Q_m)$ where $\xi (k)$ is the quasiparticle dispersion relation measured from the chemical potential.

\begin{figure}[t]
\subfigure{\includegraphics[scale=0.149]{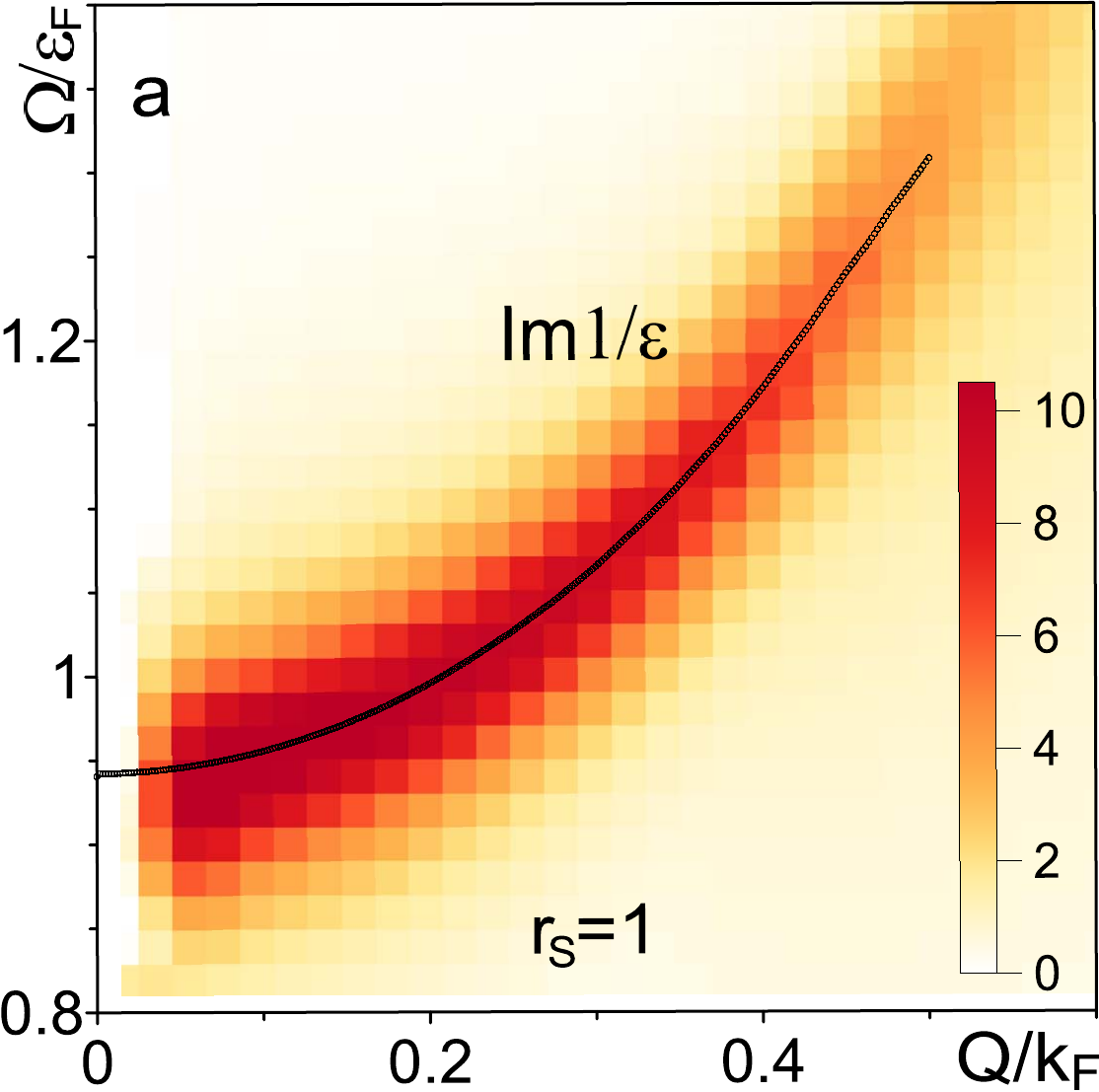}}
\subfigure{\includegraphics[scale=0.149]{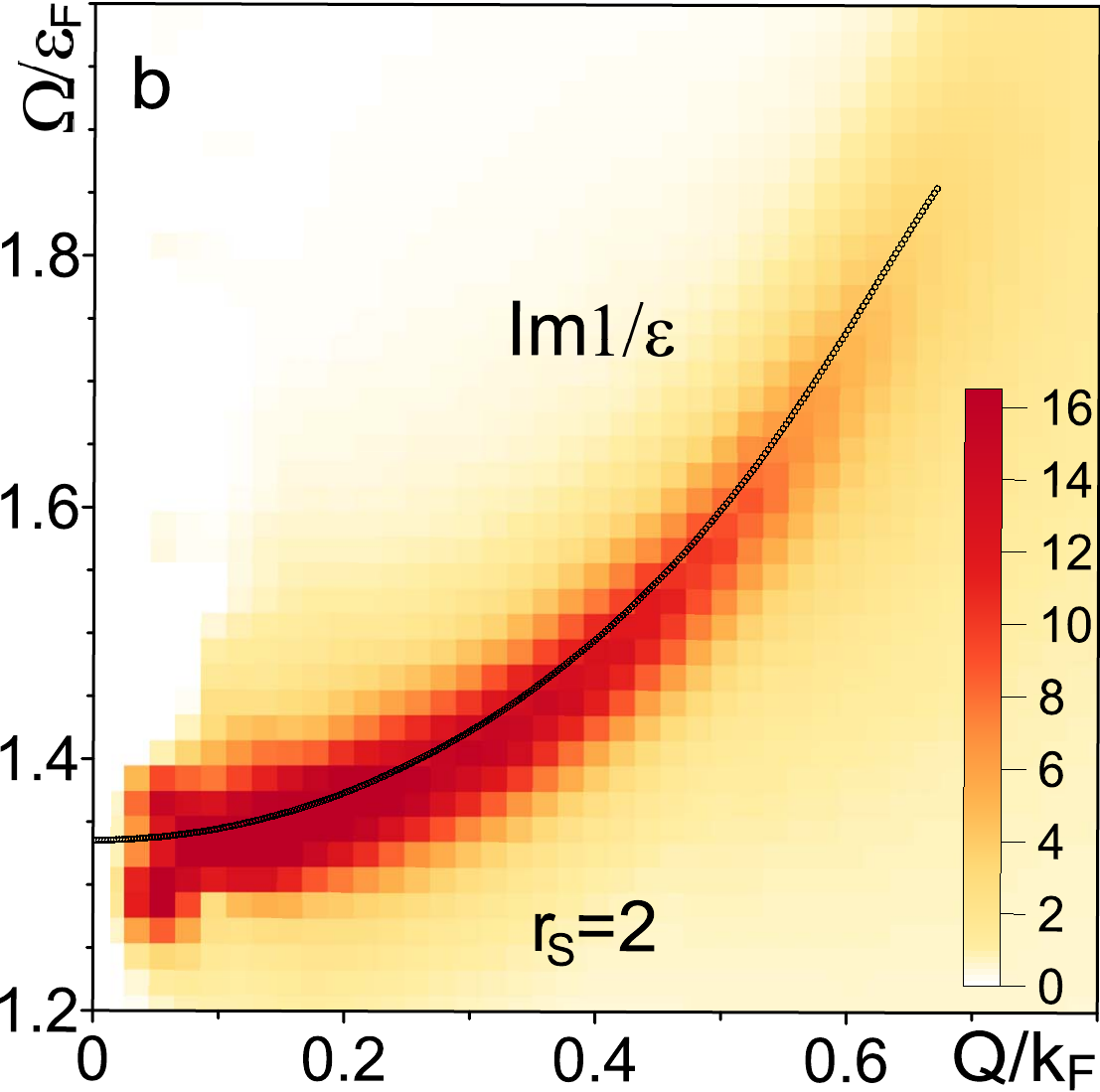}}
\subfigure{\includegraphics[scale=0.149]{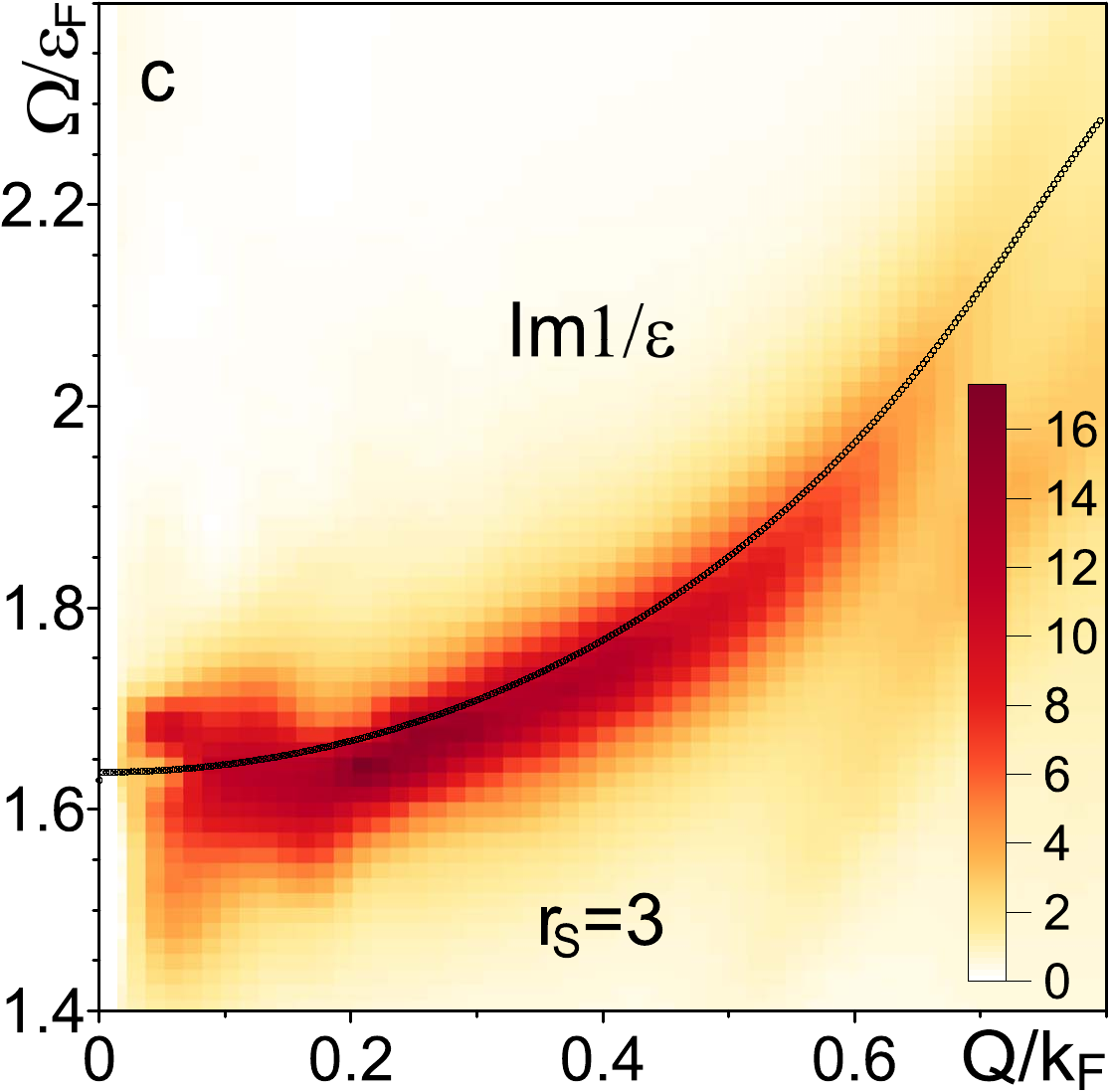}}
%\vspace{-2mm}
\caption{Loss function $\mathrm{Im} \epsilon^{-1}$ at $T/\varepsilon_F=0.1$ for $r_s = 1$ (a), $r_s=2$ (b), and $r_s=3$ (c). The plasmon dispersion in RPA is shown by small black circles. For this data set $\eta=\varepsilon_F/20$. }
\label{Fig2}
\end{figure}

The plasmon dispersion is visualized in Fig.~\ref{Fig2} showing the loss function $\mathrm{Im} \epsilon^{-1}$ in the $(Q, \Omega)$ plane. At moderate values of $r_s$ and small momenta the plasmon spectrum closely follows the RPA result at the same temperature starting from the exact hydrodynamic relation, $\omega_{\mathrm{pl}}^2(Q=0) = \Omega_{\mathrm{pl}}^2=4 \pi e^2 \rho /m$. Deviations become visible at large momenta where $r_s=2$ and $r_s=3$ loss function maxima are getting visibly lower than the RPA ones.


Broadening of the plasmon dispersion comes from decay processes into multiple particle-hole pairs. The corresponding lifetime is finite even at $T=0$~\cite{TTKP2021}. Additional contribution to broadening in Fig.~\ref{Fig2} comes from a finite value of $\eta$ in the substitution $\Omega \to \Omega + i \eta$ that provides regularization of all poles under the integrals. Ultimately, the final results need to be extrapolated to $\eta=0$, but calculations with small values of $\eta$ are progressively more expensive. A meaningful compromise is to select $\eta \ll \min \{ \varepsilon_F, T \}$ (see also Note III in the Supplemental material \cite{SM}). While choosing $\eta \sim T$ can distort the data significantly, for $\eta=\varepsilon_F/200 \ll T$ this systematic bias does not exceed $5\%$ (which is smaller than uncertainty originating from third-order expansion at $r_s > 1$). Except for Fig.~\ref{Fig2}, all data in the main text were computed with $\eta=\varepsilon_F/200$.

To obtain results with desired accuracy one has to account for high enough diagrammatic orders and the proper balance is between the systematic errors originating from the series truncation and statistical errors. Importance of high-order terms increases with $r_s$; while for $r_s=1$ calculations up to the $3$-rd order are sufficient (by observation that $3$-rd and $4$-th order results are nearly indistinguishable at $r_s=2$, see Fig.~\ref{Fig3} in the Supplemental Materials), the $r_s>1$ cases may require higher order contributions for reaching the desired accuracy (see Fig.~\ref{Fig5} below and Note IV in the Supplemental material \cite{SM}).

Third-order calculations with $\eta=\varepsilon_F/200$ take from a few days (for dielectric function curves shown in Fig.~\ref{Fig1}) to several weeks (for exchange-correlation kernel curves shown in Fig.~\ref{Fig4} below) on a $256$-core cluster. Extending these simulations to the $4$-th order is estimated to take at least a factor of ten longer (especially at high frequencies). An important algorithmic development that may reduce the computational cost would be to implement the $\eta \to 0$ limit analytically \cite{TTKP2021}.

%%%%%%%%%%%%%%%%%%%%%%%%%%%%%%%%%%%%%%%%%%%%%%%%%%%%%%%%%%%%%%%%%%%%%%%%%
\smallskip

\noindent
\textbf{Exchange-correlation kernel.} Within the TDDFT, the charge response function, $\chi(\textbf{Q},\Omega,T)$, is constructed from the non-interacting response function $\chi_{KS}$ and the exchange-correlation kernel $K_{\mathrm{xc}}(\textbf{Q},\Omega,T)$. Following Ref.~\cite{Perdue2020}), one has
\begin{equation}
\chi = \chi_{KS} / [1 - (V_0+K_{\mathrm{xc}}) \chi_{KS}],
\label{TotPi}
\end{equation}
where in jellium $\chi_{KS} = \Pi_{\mathrm{RPA}}$ is given by Eq.(\ref{RPA}) (at $T=0$ it is the Lindhard function \cite{Lindhard}). By comparing Eq.~(\ref{TotPi}) with the definition of $\chi$ through the exact polarization function, $\chi=\Pi/\epsilon$, we arrive at the definition of $K_{\mathrm{xc}}$ in terms of polarization functions
\begin{equation}
K_{\mathrm{xc}}(\textbf{Q},\Omega,T) = \Pi^{-1}_{\mathrm{RPA}}(\textbf{Q},\Omega,T) - \Pi^{-1}(\textbf{Q},\Omega,T).
\label{ExchCorr}
\end{equation}
While $\chi_{KS}$ is always straightforward to calculate, the kernel $K_{\mathrm{xc}}$ is typically approximated by a certain jellium model parametrization---its dependence on frequency is perhaps the most important challenge in the modern theory of the electron liquid~\cite{vignale}. Due to the NAC problem, conventional quantum Monte Carlo methods can not address the dynamics of realistic interacting models. In the absence of numerical inputs, the existing phenomenological approximations~\cite{Runge1984,Kohn1985,Perdew92,Perdue2020,Gunn2002} were shown to be insufficient in a number of cases \cite{Nepal}. As a result, the frequency dependence of the kernel remains largely unknown, except for known zero and infinite frequency limits.

The most prominent feature of charge response function is the plasmon resonance, see Fig.\ref{Fig3}. Its amplitude and width are controlled by the plasmon lifetime, which is finite at $(Q,T)=(0,0)$ and increases with $Q$ and $T$ \cite{TTKP2021}. In contrast, the plasmon decay into multiple electron-hole pairs is absent in the RPA and the peak in $\chi_{KS}$ is a delta function (regularized in simulations by $\eta \ne 0$). A shift in the pole position at momentum $Q \approx k_F/2$ (better seen in the $\mathrm{Im} \chi$ part) reflects deviations in the plasmon dispersion relation from the RPA prediction (see also Fig.~\ref{Fig2}).

\begin{figure}[t]
%\vspace{-2mm}
\subfigure{\includegraphics[scale=0.25]{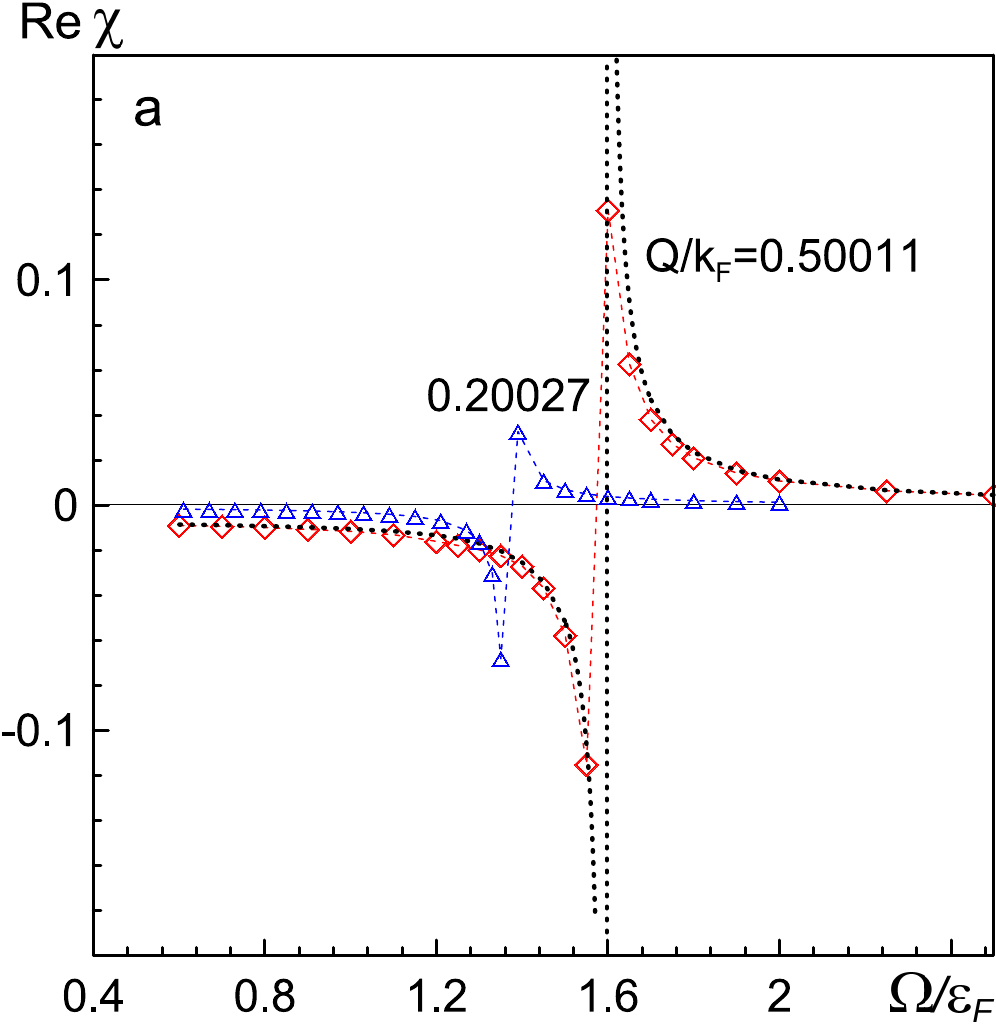}}
\subfigure{\includegraphics[scale=0.25]{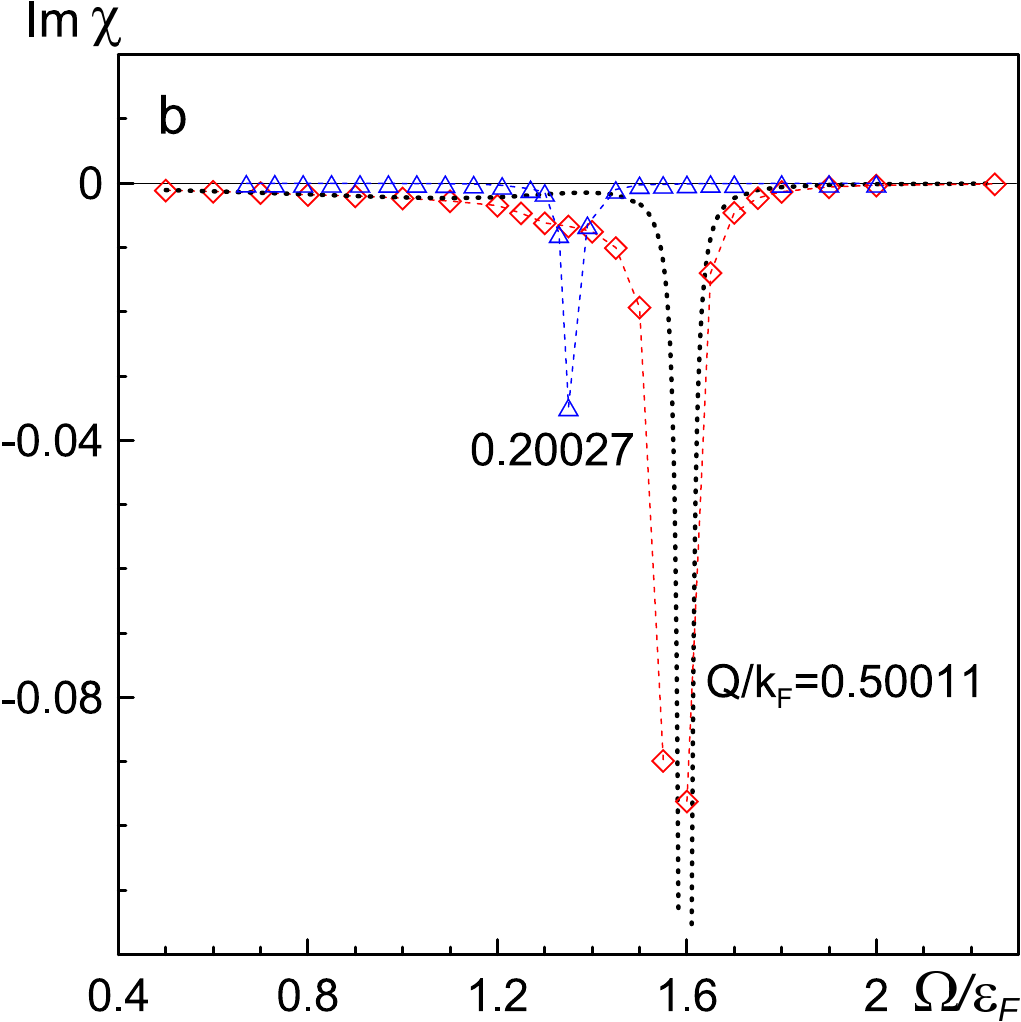}}
%\vspace{-3mm}
\caption{Real (a) and imaginary (b) parts of $\chi$ as functions of $\Omega$ at $T/\varepsilon_F=0.1$ for $r_s = 2$ and momenta $Q/k_F=0.20027$, $0.50011$. Simulation results are shown with red and blue curves with symbols. Black dotted curves: $\eta$-dependent RPA results for $Q/k_F=0.50011$ truncated at the figure scale. Errors are within the symbol sizes. }
\label{Fig3}
\end{figure}
\begin{figure}[h]
\subfigure{\includegraphics[scale=0.138]{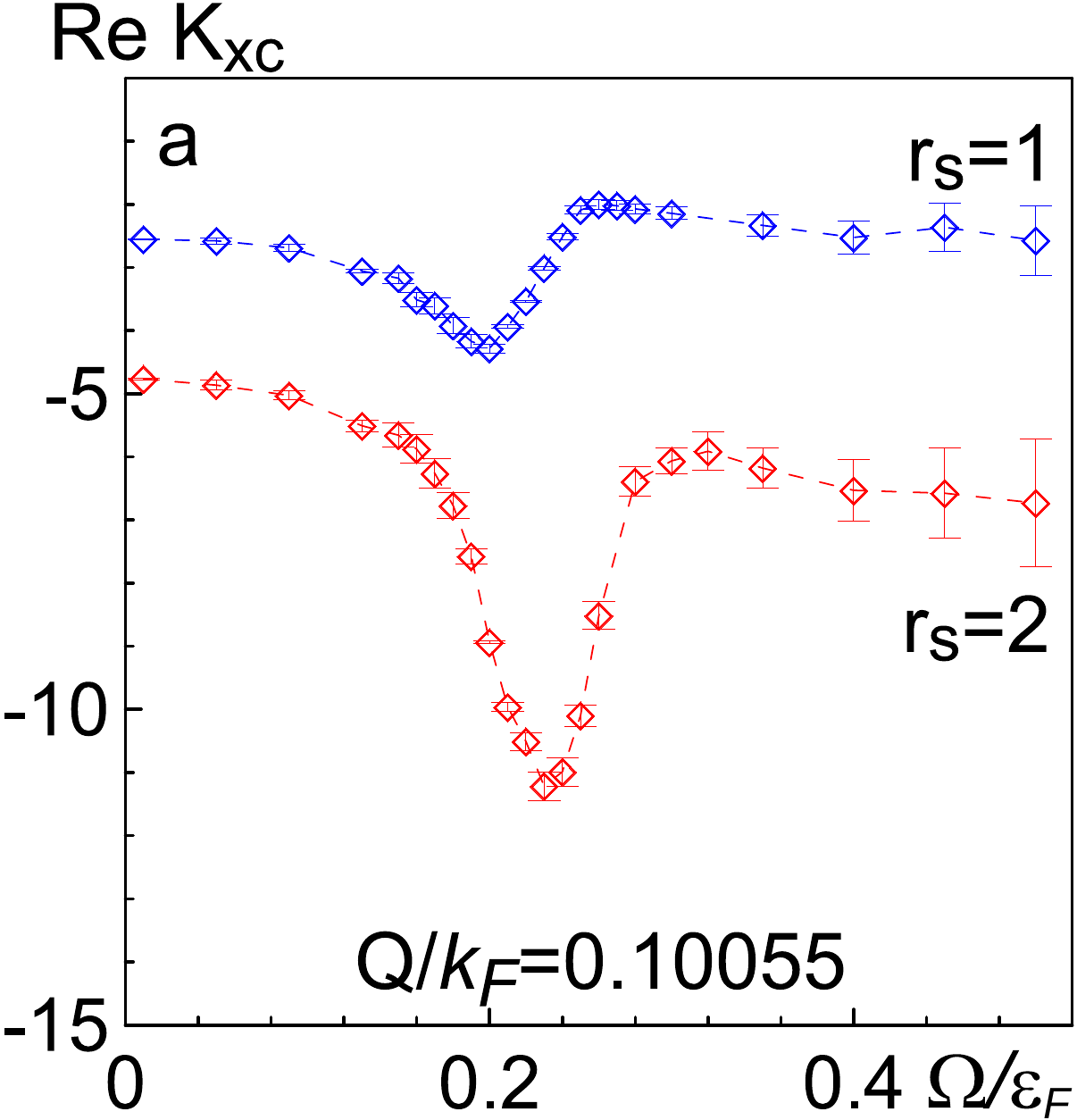}}
\subfigure{\includegraphics[scale=0.138]{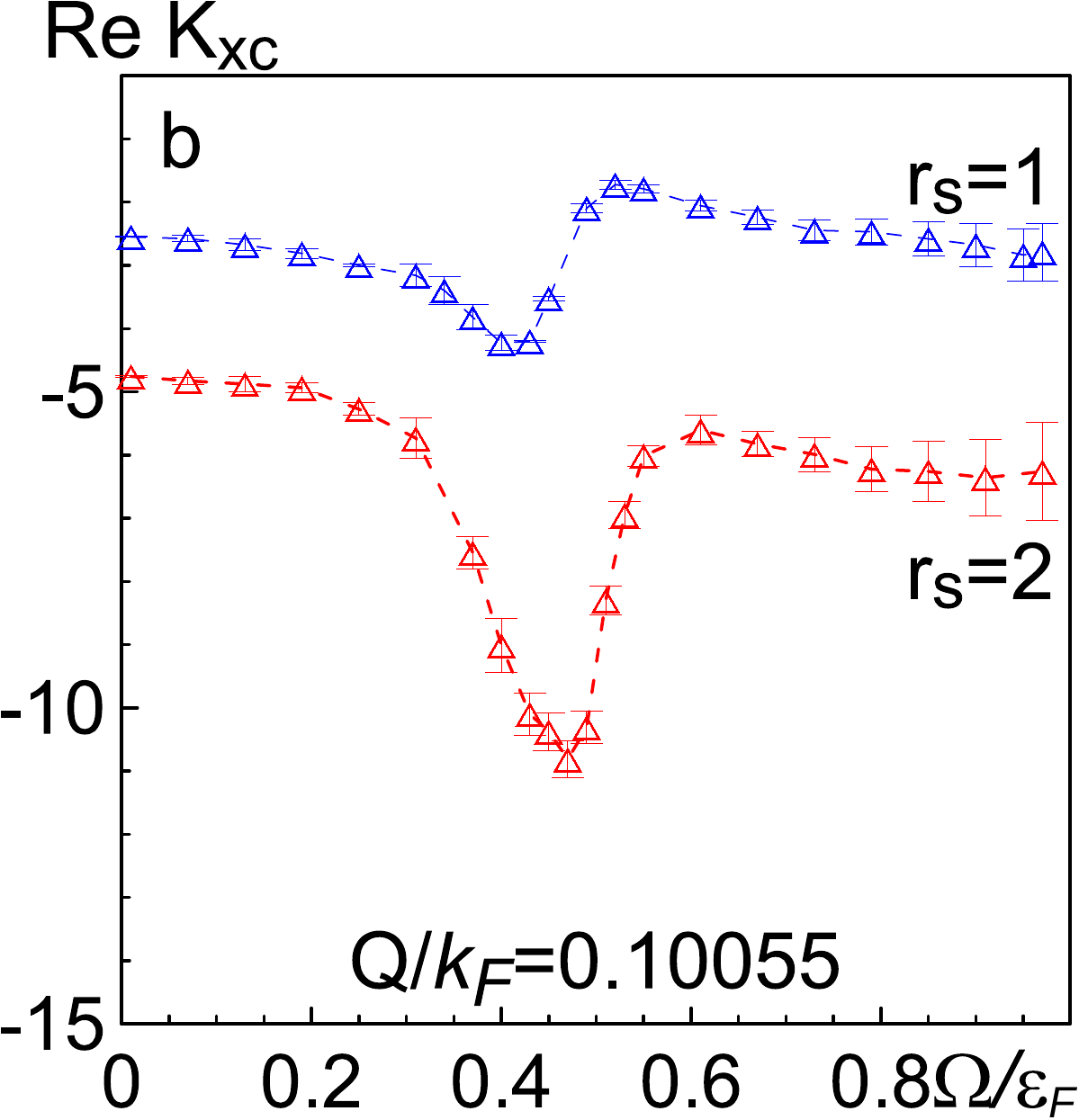}}
\subfigure{\includegraphics[scale=0.138]{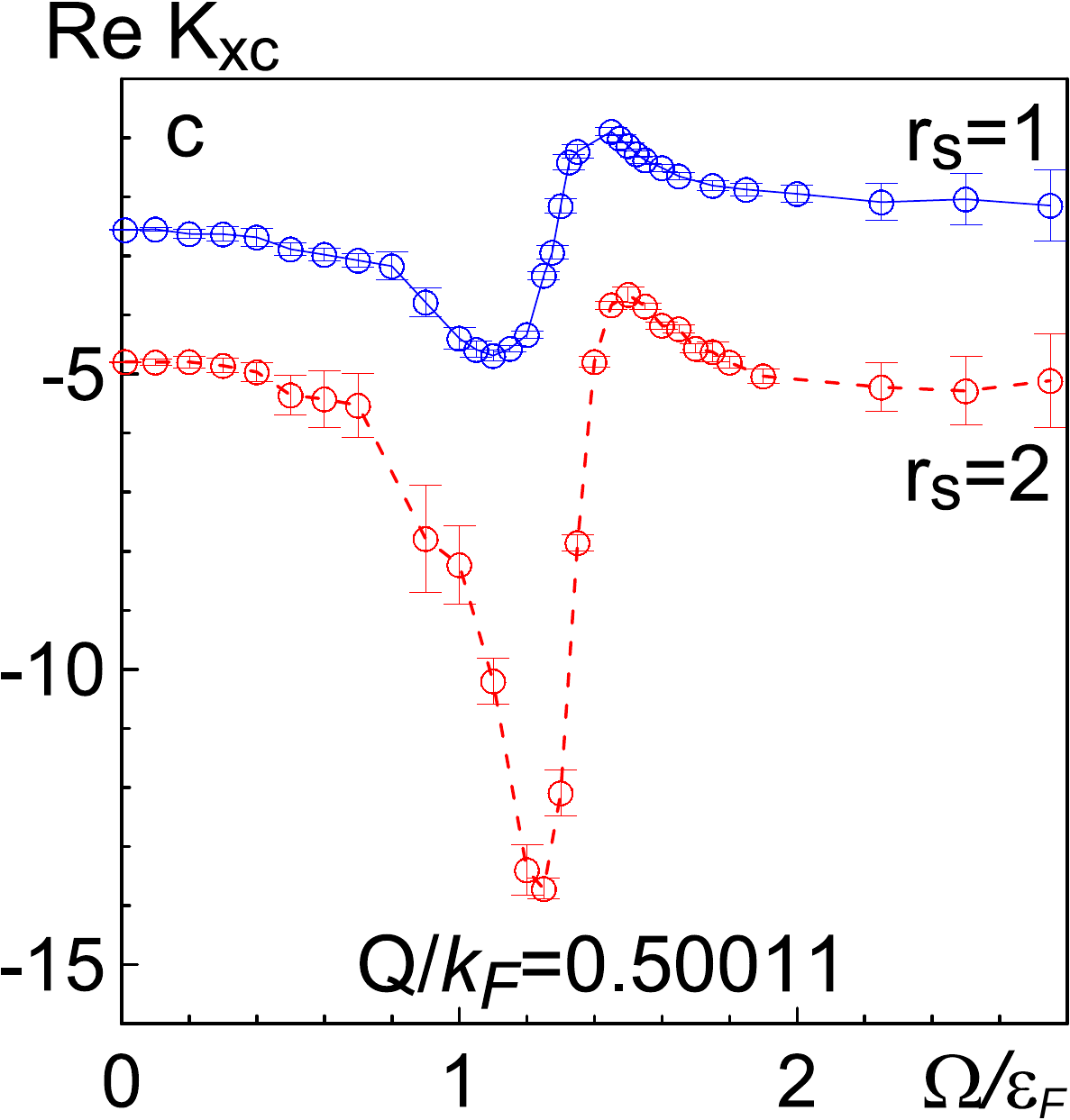}}
\subfigure{\includegraphics[scale=0.138]{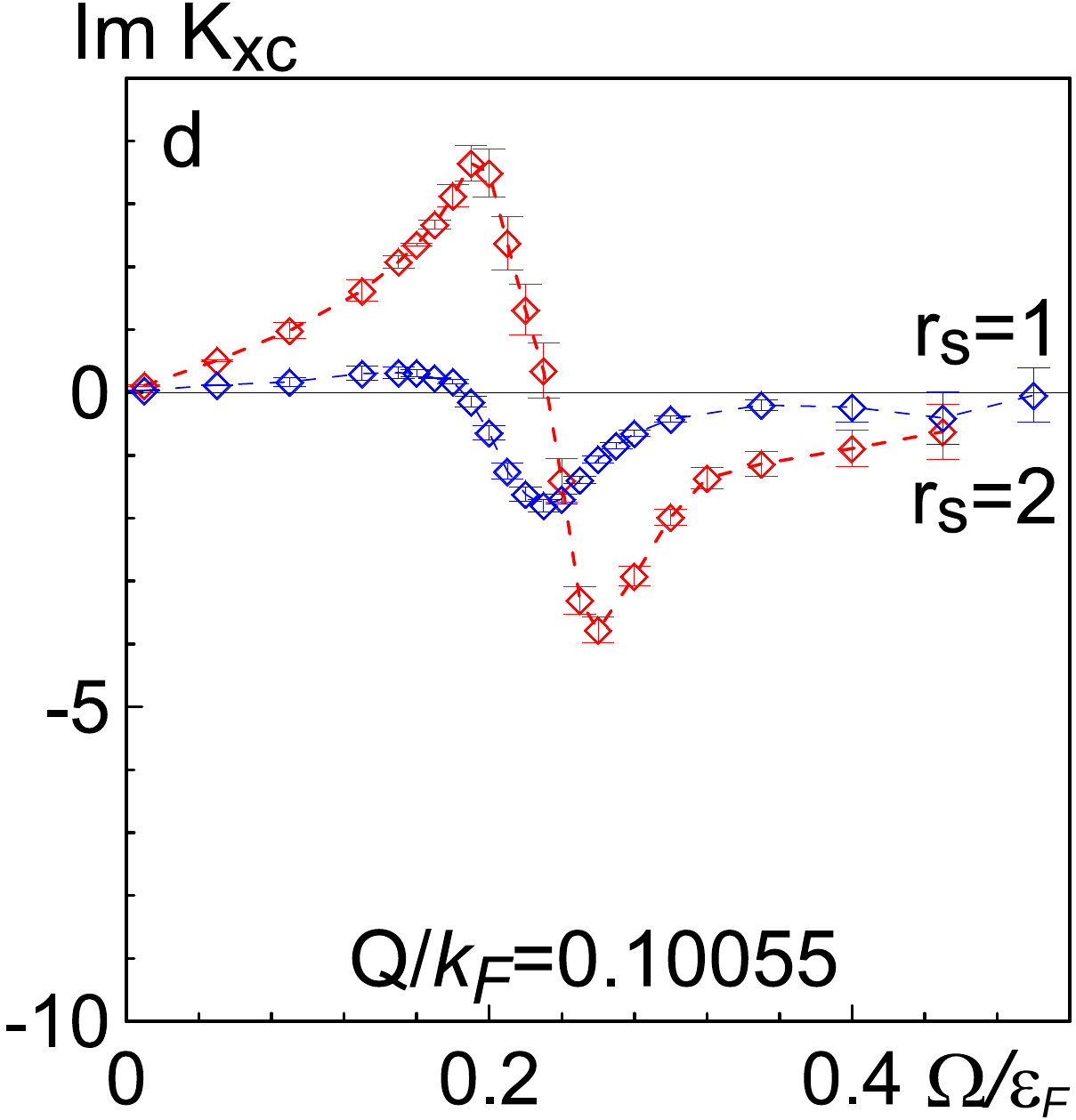}}
\subfigure{\includegraphics[scale=0.138]{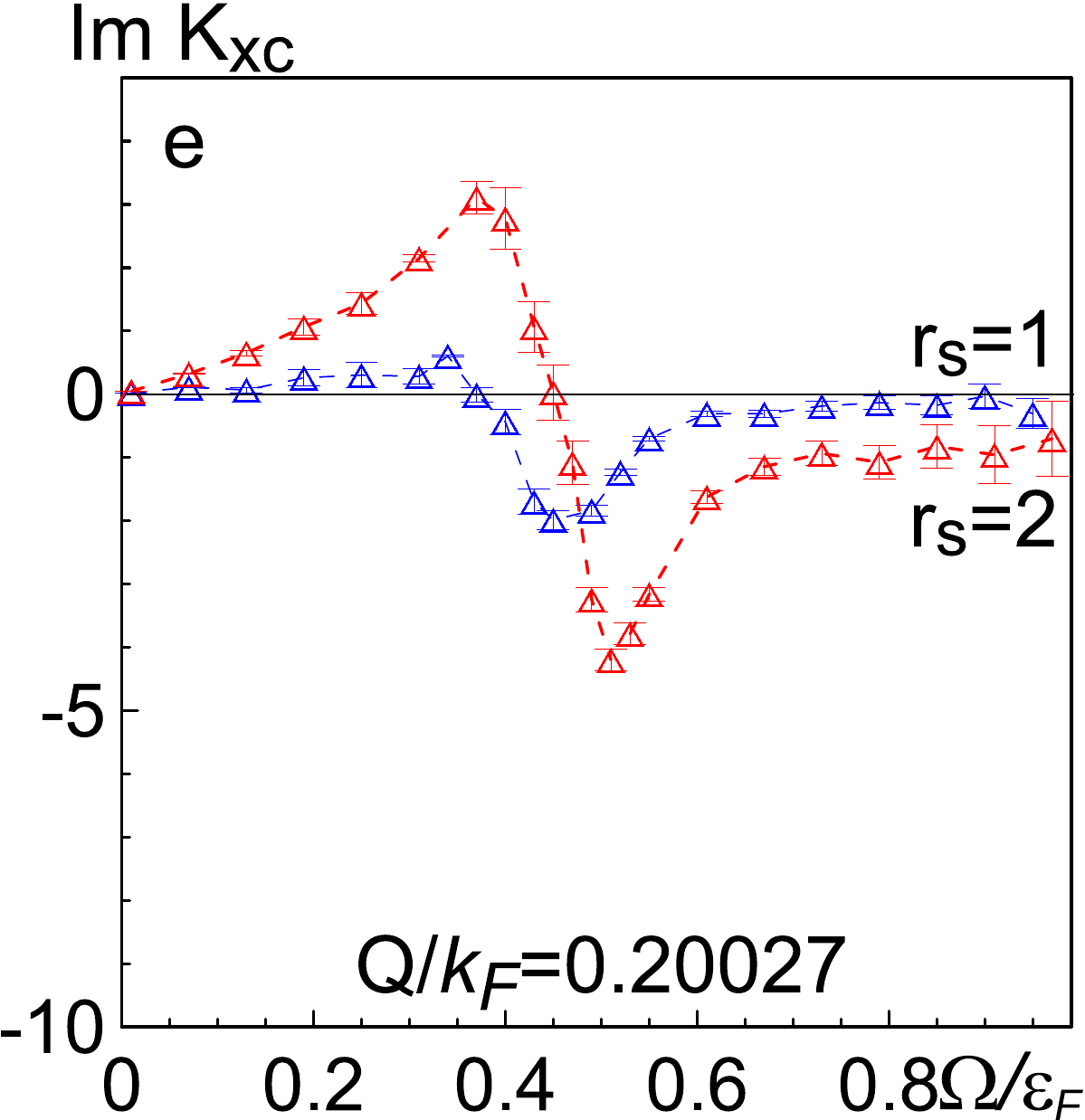}}
\subfigure{\includegraphics[scale=0.138]{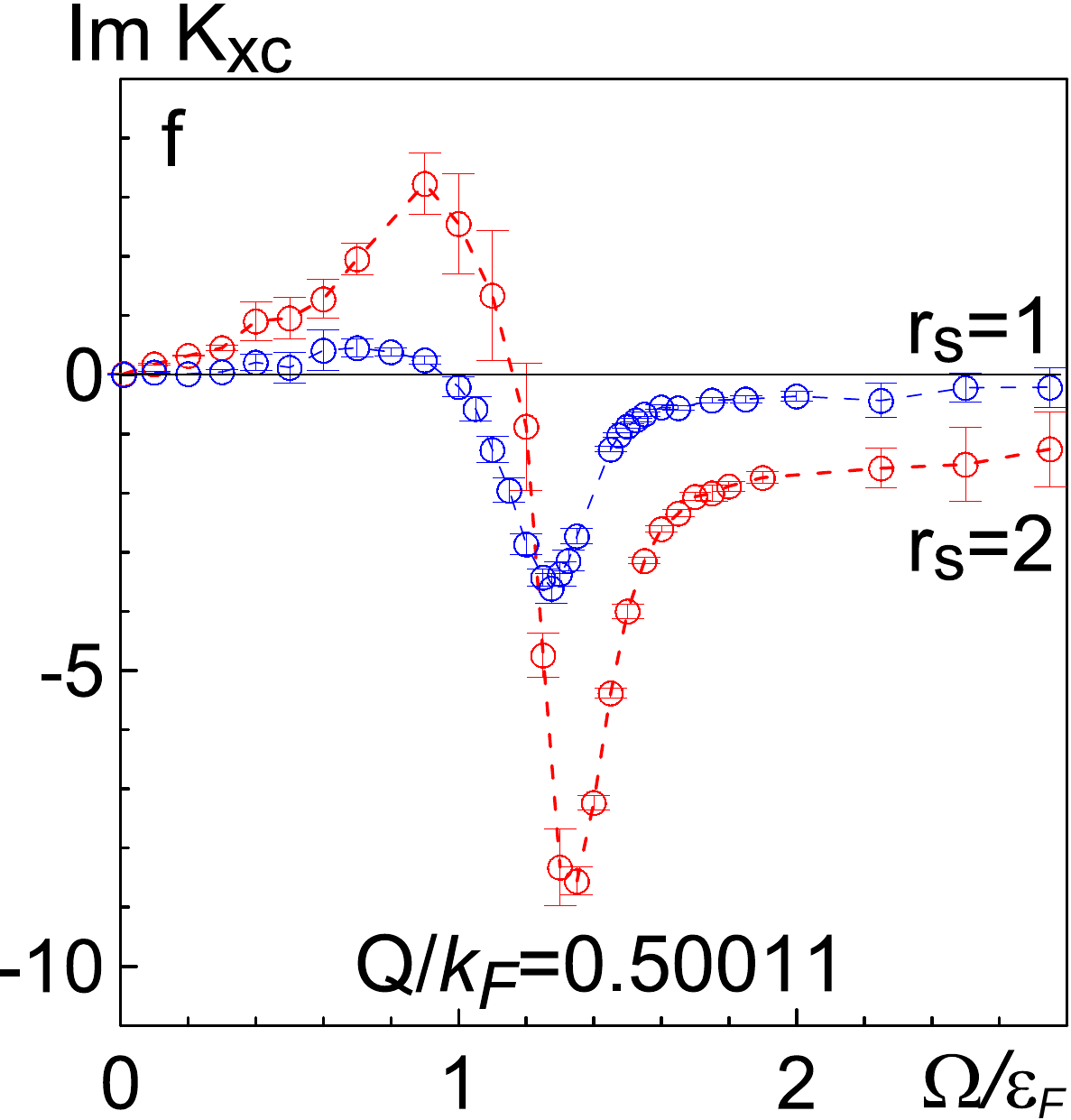}}
%\vspace{-3mm}
\caption{Real (a,b,c) and imaginary (d,e,f) parts of the exchange-correlation kernel $K_{\mathrm{xc}}$ in jellium as functions of frequency at $T/\varepsilon_F=0.1$, $r_s = 1$ and $2$, and several values of momentum $Q$.}
\label{Fig4}
\end{figure}

Our finite-$T$ simulations of the exchange-correlation kernel $K_{xc}(\Omega/\varepsilon_F)$ are shown in Fig.~\ref{Fig4}. They are based on the exact relation (\ref{ExchCorr}) and simulated on a relatively sparse $\{Q,\Omega \}$-grid with momenta $Q \lesssim k_F/2$ with the goal of demonstrating the feasibility of the technique. Proper tabulation of the kernel on a dense grid for practical TDDFT applications goes beyond the scope of present work (and requires substantial increase in computational resources).

$K_{xc}(\Omega/\varepsilon_F)$ curves feature two prominent extrema around $\Omega \sim v_F Q$, which grow in amplitude with $Q$ and $r_s$, and have been previously missed by phenomenological modeling of $K_{xc}$. They are related to multiple crossings between the high-order (3rd-order in Fig.~\ref{Fig4}) and RPA polarization functions (see also Note V in the Supplemental material \cite{SM}) determined by properties of the $e-h$ continuum. Unlike RPA, high-order results include contributions from multiple excitation processes in addition to renormalization of the single particle dispersion and $Z$-factor. We note that the imaginary part of $K_{\mathrm{xc}}(\Omega)$ is positive at small frequencies and goes negative only beyond the frequency  $\Omega > v_F Q$, hence $K_{xc}$ is not causal as is frequently assumed.

\begin{figure}[h]
%\vspace{-2mm}
\includegraphics[scale=0.4]{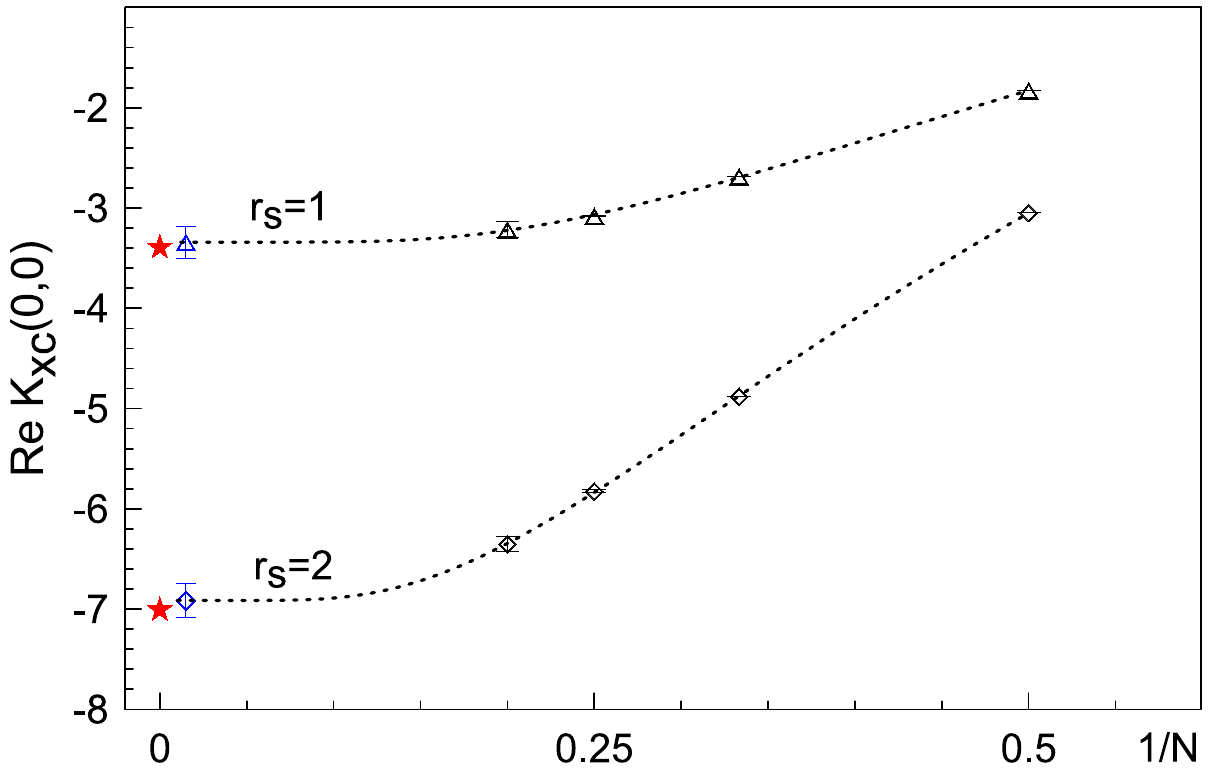}
%\vspace{-2mm}
\caption{Real parts of the exchange-correlation kernel $K_{\mathrm{xc}}(Q=0, \Omega=0)$ at $T/\varepsilon_F=0.1$ as functions of the inverse diagrammatic order $N$ for $r_s = 1$ (triangles) and $r_s=2$ (diamonds). Exponential fits $a+be^{-cN}$ (black dotted lines) were used to perform extrapolation towards an infinite diagrammatic order limit shown by blue symbols. Red stars: static $\mathrm{Re}K_{\mathrm{xc}}(Q=0)$ from \cite{Perdue2020} (in our units).}
\label{Fig5}
\end{figure}

The large frequency limit of $K_{\mathrm{xc}}$ is known from the exchange-correlation energies of the model~\cite{Kugler1975}, but this asymptotic regime has not been reached in our simulations because the difference between the exact and RPA response functions becomes vanishingly small at high frequency
while both quantities tend to zero, leading to strongly amplified numerical noise in $K_{\mathrm{xc}}$ data, and consequently large error-bars. However, the important low and intermediate frequency parts of the kernel at finite momentum are not masked by noise. It is also evident that for $\Omega \ll \epsilon_F$ and $Q\ll k_F$ these curves are self-similar functions that depend only on the $\Omega / v_F Q$ ratio, i.e., minima and maxima shift to smaller and smaller frequencies when $Q\to 0$ and $\mathrm{Re} K_{\mathrm{xc}}(\textbf{Q}, \Omega=0)$ saturates to its finite $Q=0$ limit---at $T=0$ it is determined by the derivatives of the exchange-correlation energy with respect to density \cite{Kugler1975,Kohn1985,Perdue2020}.

Our data for $\mathrm{Re}K_{\mathrm{xc}}(Q,\Omega=0,T)$ largely agree with, but numerically do not precisely match the values presented in \cite{Kohn1985,Perdue2020} on the basis of ground state calculations (after conversion to the same units). This is mainly the finite diagrammatic order effect. A few percent contribution to $\Pi$ from higher order diagrams results in a much larger effect for the difference $\Pi - \Pi_{\mathrm{RPA}}$ determining the kernel (see the $\Omega \to 0$ limit in Fig.~\ref{Fig1}). In Fig.~\ref{Fig5} we show that results for $\mathrm{Re}K_{\mathrm{xc}}(Q=0,\Omega=0,T)$ computed up to fifth order and extrapolated to an infinite order limit do match static ground state answers within the errorbars.

%%%%%%%%%%%%%%%%%%%%%%%%%%%%%%%%%%
\smallskip

\noindent \textbf{Conclusions.} By implementing the algorithmic Matsubara integration within the diagrammatic Monte Carlo approach we formulated a technique for accurate calculations of dynamic response in the homogeneous electron gas at finite temperature. It works directly in the real-frequency domain and eliminates the need for the infamous numerical analytic continuation---the long-standing obstacle for the accurate theoretical description of experimentally relevant observables.

We computed the exchange-correlation kernel of the homogeneous electron gas by a controlled method for the first time, and revealed unexpected features in its frequency dependence, which should spark the development of better kernels for the time-dependent density functional theory both at zero and finite temperature.

\smallskip

\noindent \textbf{Acknowledgements.} I.S.T. and N.V.P. thank support from DOE DE-SC0023141. K. Ch. 
is thankful for support from the Simons Collaboration on the Many Electron Problem and Flatiron Institute; J.P.F.L. thanks support of the Natural Sciences and Engineering Research Council of Canada (NSERC) RGPIN-2017-04253; K.H. thanks support from NSF DMR-1709229;

\end{document}

\title{Supplemental material for "Dynamic Response of an Electron Gas: Towards the Exact Exchange-Correlation Kernel"}

\author{James P. F. LeBlanc}
\author{Kun Chen}
\author{Kristjan Haule}
\author{Nikolay V. Prokof'ev}
\author{Igor S. Tupitsyn}

\maketitle

%%%%%%%%%%%%%%%%%%%%%%%%%%%%%%%%%%%%%%% TITLE %%%%%%%%%%%%%%%%%%%%%%%%%%%%%%%%%%%%

%\smallskip

\begin{center}
\textbf{Note I: Diagrammatic groups}
\end{center}

\begin{figure}[tbh]
%\vspace{-2mm}
\includegraphics[scale=0.35]{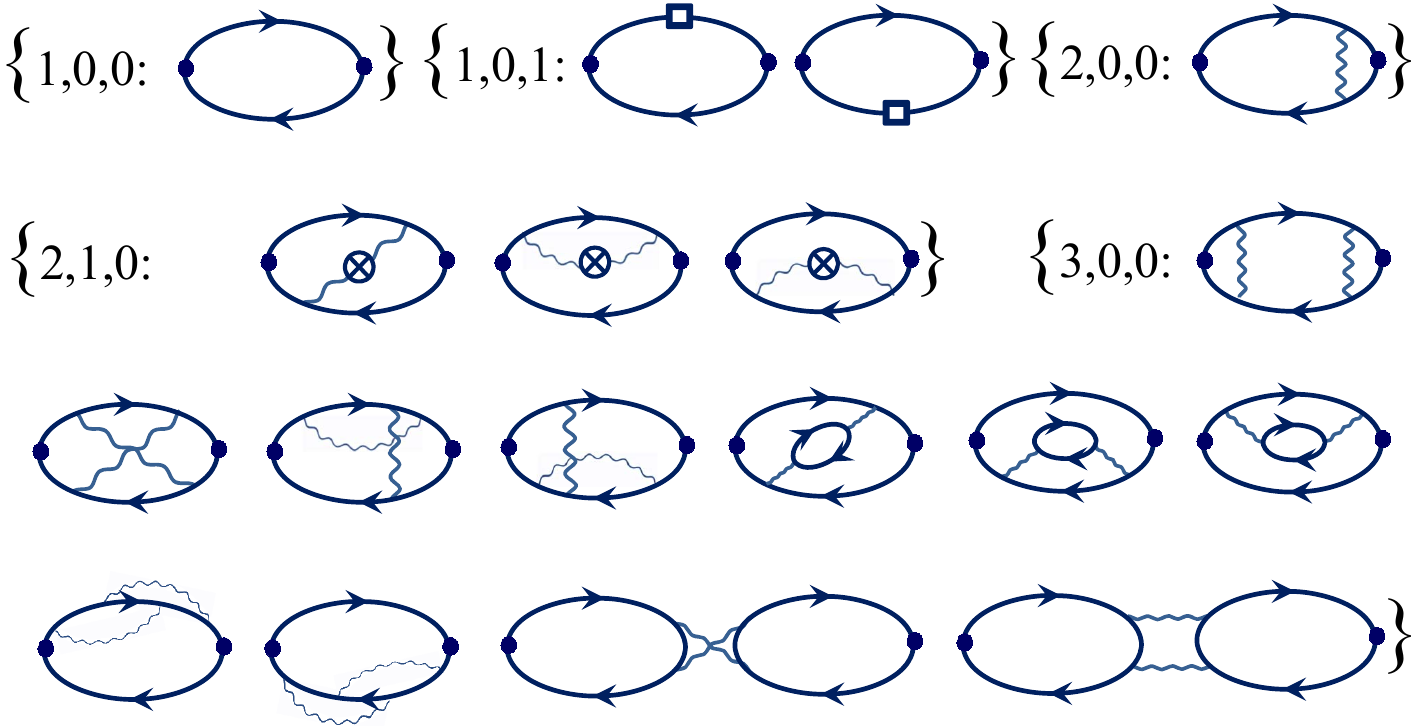}
%\vspace{-2mm}
\caption{Diagrams contributing to the irreducible polarization function. Open squares in the group $(1,0,1)$ and crossed circles in the group $(2,1,0)$ stand for self-energy and polarization counter-terms, respectively}
\label{SFig1}
\end{figure}

Figure~\ref{SFig1} shows diagrams contributing to the irreducible polarization function $\Pi({\bf Q},\Omega,T)$ up to the third order inclusively (below we use the terminology introduced in Ref.~\cite{Chen2019}).
Diagrams are separated into groups depending on the number of counter-terms: groups $100$ ($\{a,b,c\}=\{1,0,0\}$), $200$ ($\{a,b,c\}=\{2,0,0\}$), and $300$ ($\{a,b,c\}=\{3,0,0\}$) are the standard diagrams for polarization function defining the first, second, and third expansion orders. The $101$ ($\{a,b,c\}=\{1,0,1\}$) and $210$ ($\{a,b,c\}=\{2,1,0\}$) groups contain the self-energy and polarization counter-terms---both these groups are of the third order as are diagrams in the $300$ group. All other groups, not listed in Fig.~\ref{SFig1}, belong to higher expansion orders.

The unique $100$ and $200$ diagrams correspond to the polarization bubble and the leading vertex correction respectively, see Fig.~\ref{SFig1}. In the Hartree-Fock basis there are $11$ diagrams belonging to the $300$ group (the group with two interaction lines and no counter-terms) and $5$ diagrams in the $101$ and $210$ groups (counter-terms). There are $132$ diagrams of forth-order in groups $201$, $220$, $310$, and $400$; the $400$ group is characterized by three interaction lines and contains $95$ terms. After algorithmic Matsubara integration the total number of terms to be summed up and integrated using Monte Carlo sampling at 3d-order is $720$. At 4th-order one has to deal with $6659$ terms generated by the AMI procedure; most terms have multiple sharp sign-alternating singular contributions.

%By the ``3d-order result'' in the main text we mean the sum of all diagrams up to the third order %(those listed in Fig.~\ref{SFig1}).

\begin{center}
\textbf{Note II: Analytic Wick rotation to the real frequency axis}
\end{center}

As noticed in Ref.~\cite{LeBlanc2019}, sums over internal  Matsubara frequencies can be performed
analytically if the series are expressed in terms of the Green's function with a simple pole structure and frequency independent interaction. In this case, the frequency dependent part of all diagrammatic contributions is a product of simple poles, $\prod_s G (i\omega_s, \mathbf{k}_s) \equiv \prod_s  [i\omega_s - \varepsilon(\mathbf{k}_s)]^{-1}$ (the spin index is omitted for brevity), where $\omega_s$ are fermionic Matsubara frequencies. All sums over Matsubara frequencies for such products can be completed analytically with the help of the Cauchy formula
\begin{equation}
T \sum_n \left( \prod_{j=1}^{M} \frac{1}{i\omega_n -a_j}  \right) = \sum_{j=1}^{M} n_j
\left( \prod_{s\ne j}^{M} \frac{1}{a_j-a_s}  \right) ,
\label{Cauchy}
\end{equation}
where $n_j \equiv n(a_j)=[e^{a_j/T}+1]^{-1}$ is the Fermi-Dirac function. Bose-Einstein function is related to $n(a)$ by $N(a) =-n(a + i \pi T)$; thus, summation over bosonic frequencies $\omega_m$ is included in Eq.~(\ref{Cauchy}) by the transformation $i \omega_m - a  =  i \omega_{m+1/2} - a'$ with $a'=a + i \pi T$. This procedure can be automated for any order diagram and since the final expression is an analytic function of external Matsubara frequency, the Wick rotation to the real-frequency axis, $i\omega_e \to \Omega + i\eta$, is trivial. The integration/summation over the remaining internal variables (momenta, spins, etc) is then performed using standard MC methods to produce results directly on the real-frequency axis and eliminating the need for NAC.

Since the self-consistent Hartree-Fock solution preserves the simple pole structure of $G$ and Yukawa potential is frequency independent, this method, or AMI (see Ref.~\cite{LeBlanc2019} for more details), is applicable to the jellium. Most importantly, the AMI automates the generation of all terms in Eq.~(\ref{Cauchy}).

%\begin{center}
%\textbf{Polarization counter-terms}
%\end{center}
%
%Since the diagrammatic expansion in terms of the bare Coulomb potential $V_0$ is ill-defined (diagrams %with bubble insertions produce divergent contributions), we reformulate the action using shifted action %tools \cite{ShiftAct}. First, $V_0$ is replaced with the Yukawa potential, $Y=4 \pi e^2/(q^2 + %\kappa^2)$. To understand how compensation for an arbitrary screening momentum $\kappa$ works, consider %an effective Coulomb potential, $U = V_0/[1-V_0\Pi]$, and rewrite it identically as $U \equiv %Y/[1-Y(\Pi + \kappa^2/4 \pi e^2)]$. In other words, if the bare Coulomb potential $V_0(q)$ is replaced %with $Y(q)$, than the diagrammatic expansion in powers of $Y$ should be augmented with counter-terms %defined by $\kappa^2/4 \pi e^2$ and to be added to all bubble insertions (for details see %Ref.~\cite{Chen2019}).

\begin{center}
\textbf{Note III: Effect of regularization parameter $\eta$}
\end{center}

\begin{figure}[h]
\subfigure{\includegraphics[scale=0.245]{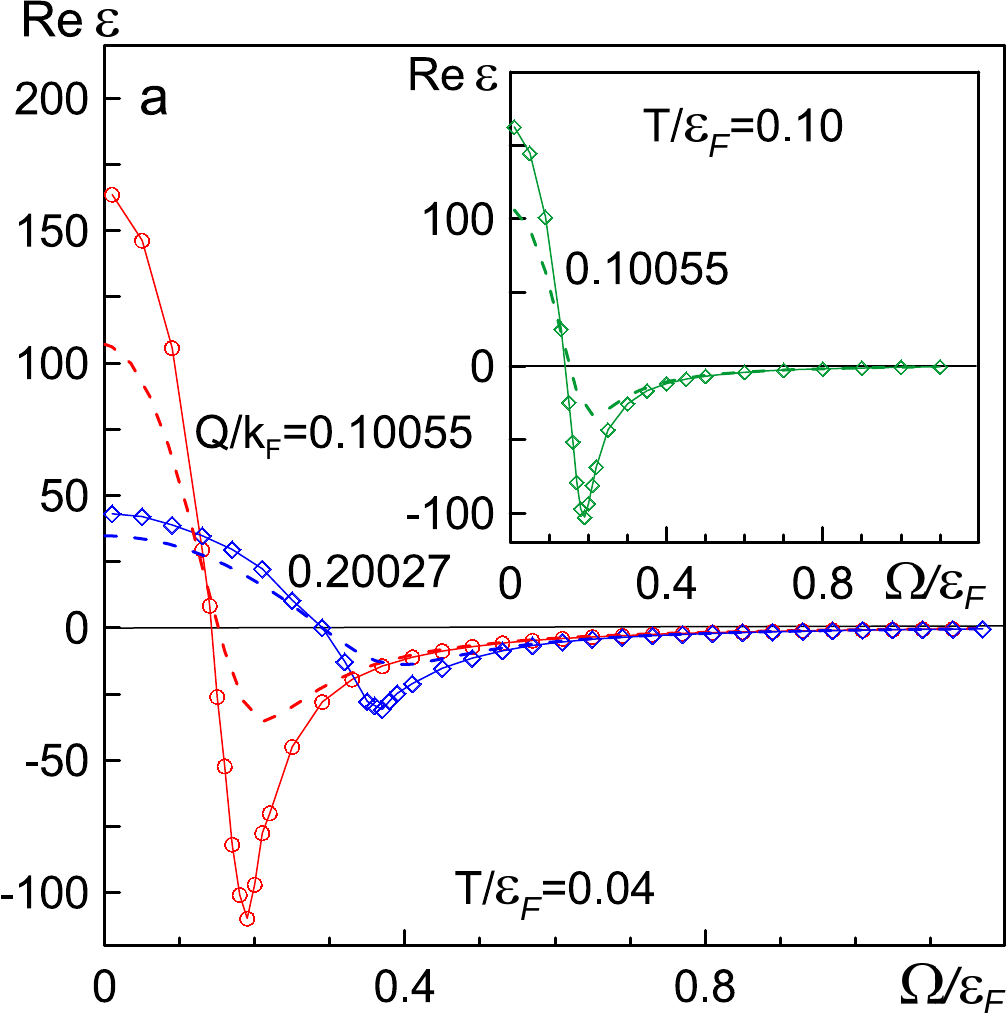}}
\subfigure{\includegraphics[scale=0.245]{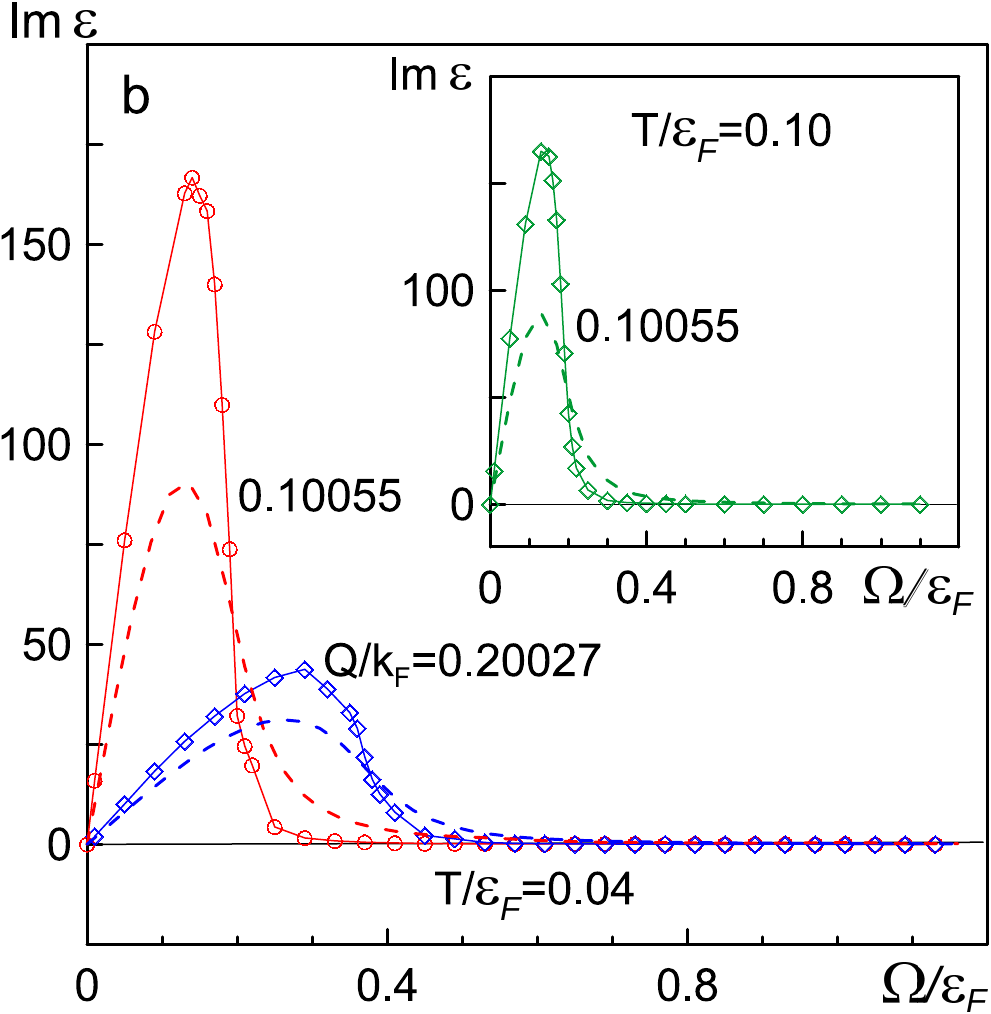}}
%\vspace{-2mm}
\caption{Real (a) and imaginary (b) parts of $\epsilon$ as functions of frequency for several momentum values at $T/\varepsilon_F=0.04$ and $r_s = 2$. Solid curves with symbols: $\eta=\varepsilon_F/200$. Dashed lines: $\eta=\varepsilon_F/20$. Insets in both figures show data for the same parameters for $Q/k_F = 0.10055$ and $T/\varepsilon_F=0.1$. Errors are within the symbol sizes. }
\label{SFig2}
\end{figure}

In Fig.~\ref{SFig2} we compare the dielectric functions $\epsilon$ obtained for two values of $\eta$ that differ by an order of magnitude.

The most sensitive to the choice of $\eta$ are the low and moderate frequency and momentum results. At high frequency and momentum the effect of finite $\eta$ is less pronounced. Large values of $\eta$ tend to suppress the amplitude and broaden non-monotonic features. Systematic errors related to $\eta$ can be eliminated by extrapolating to $\eta=0$. In practice, it is sufficient to keep them smaller than other systematic and statistic errors by taking $\eta << \min \{ \varepsilon_F, T\}$. We find that for $T/\varepsilon_F =0.1,\, 0.04$ working with $\eta \leq \varepsilon/200$ is enough to produce results with $\sim 5\%$ accuracy.

Note also that finite $\eta$ may be attributed to the realistic finite frequency resolution of the experimental technique used to probe the charge response, and as such is a convenient tool for simulating experimental signals.

\begin{center}
\textbf{Note IV: Effect of $4$-th order corrections}
\end{center}

\begin{figure}[h]
\subfigure{\includegraphics[scale=0.25]{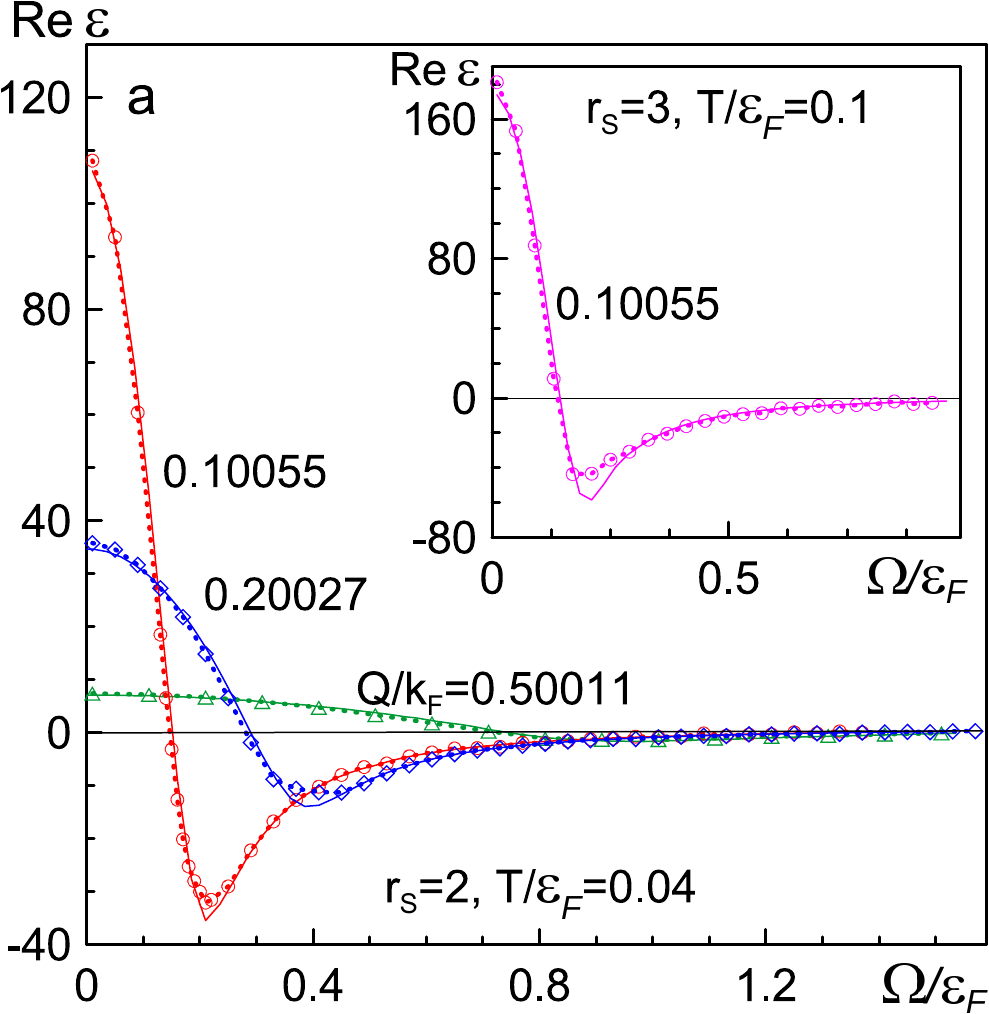}}
\subfigure{\includegraphics[scale=0.25]{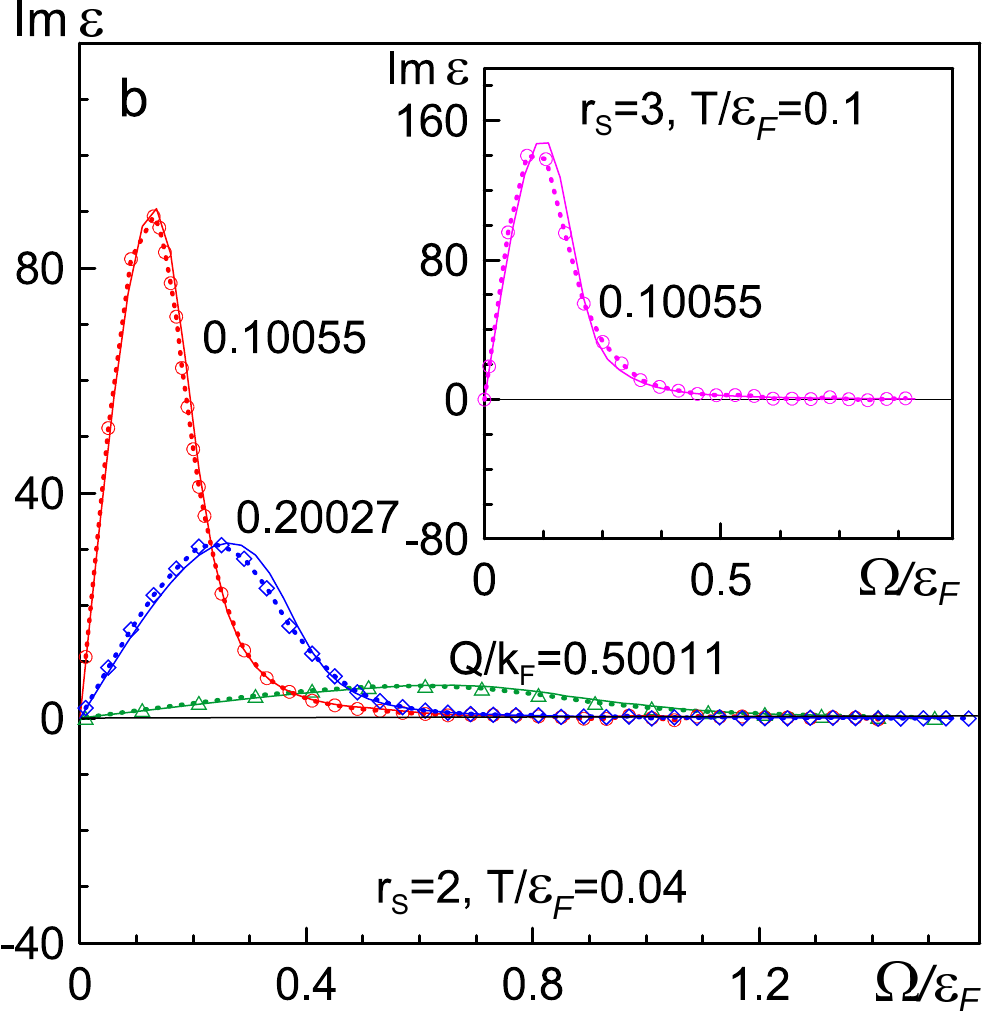}}
%\vspace{-2mm}
\caption{Real (a) and imaginary (b) parts of $\epsilon$ as functions of frequency for several momenta at $T/\varepsilon_F=0.04$ and $r_s = 2$. Solid curves: $3$rd-order results. Dotted curves with symbols: $4$th-order results. Insets in both figures show a similar comparison for $Q/k_F = 0.10055$, $T/\varepsilon_F=0.1$, $r_s=3$. $\eta=\varepsilon_F/20$.}
\label{SFig3}
\end{figure}

In Fig.~\ref{SFig3} we address the issue of higher-order corrections by comparing results for the dielectric function computed at $r_s=2$ and $3$ within the $3$-rd and $4$-th order schemes at $\eta=\varepsilon_F/20$.

Unfortunately, the full-scale $4$th-order calculations, given currently available to us computational resources, can only be done for this relatively large value of $\eta$. We observe that for such $\eta$ the $4$th-order diagrams barely change the answers and the corresponding systematic errors are comparable to statistical ones even for $r_s=3$. However, for $\eta = \varepsilon_F/200$ the $4$-th order result obtained at $r_s=2$ and $T/\varepsilon_F=0.1$ in the limit $\{ Q, \Omega \} \to 0$ already exceeds the $3$-rd order one by $\sim 5\%$ (see also Fig. 5 in the main text for the effect of higher orders on the exchange-correlation kernel).

\begin{center}
\textbf{Note V: Polarization function}
\end{center}

\begin{figure}[h]
\subfigure{\includegraphics[scale=0.245]{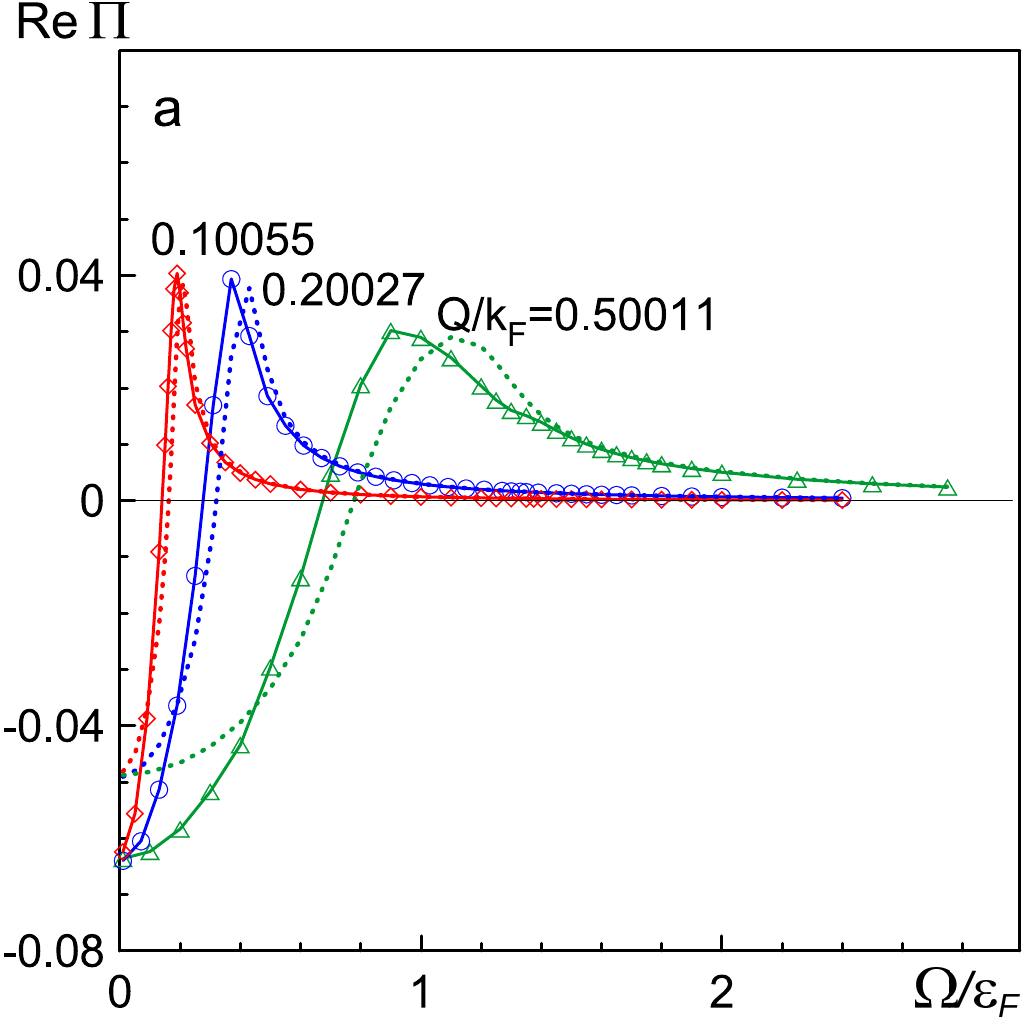}}
\subfigure{\includegraphics[scale=0.245]{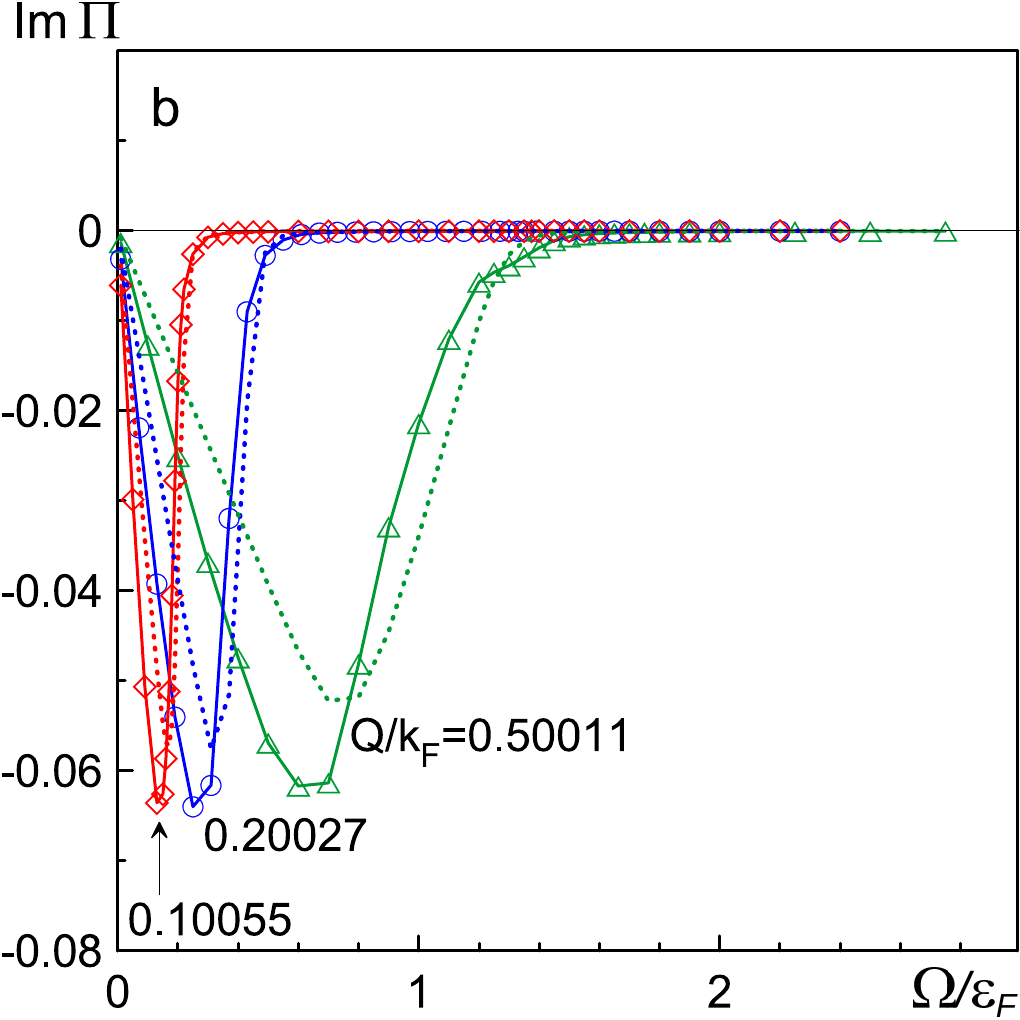}}
%\vspace{-2mm}
\caption{Real (a) and imaginary (b) parts of $\Pi$ as functions of frequency $\Omega$ at $T/\varepsilon_F=0.1$ and $r_s = 2$.
$3$rd-order results are shown by solid curves with symbols.
Dotted curves are obtained with the $(T, \eta)$-dependent RPA with $\eta=\varepsilon_F/200$. Errors are within the symbol sizes. }
\label{SFig4}
\end{figure}

Simulations of the irreducible polarization $\Pi$ determine the dielectric function, $\epsilon = 1 - V_0 \Pi$, and the charge response, $\chi = \Pi / \epsilon$. In Fig.~\ref{SFig4} we compare the 3rd-order results for $\Pi$ with the random phase approximation (RPA) answer, $\Pi_{\mathrm{RPA}}$, for $r_s=2$ (see also main text). The extrema are located at $\Omega \sim v_F Q$ and are caused by electron-hole excitations.

Real parts of $\Pi$ and $\Pi_{\mathrm{RPA}}$ are even functions of frequency
and in the asymptotic $\Omega \to \infty$ limit $|\mathrm{Re}\Pi-\mathrm{Re}\Pi_{\mathrm{RPA}}| \sim \Omega^{-4}$ (or, equivalently, $\mathrm{Re}\Pi_{\mathrm{RPA}}/\mathrm{Re}\Pi - 1 \sim \alpha(Q,T) \Omega^{-2}$). For $Q/k_F \ll 1$ this result immediately follows from the hydrodynamic relation
$\mathrm{Re} \Pi (Q \to 0) \to \rho Q^2 / m\Omega^2 $, implying that the leading $\propto \Omega^{-2}$ term is the same in $\mathrm{Re} \Pi$ and $\mathrm{Re} \Pi_{\mathrm{RPA}}$.

The difference between the imaginary parts of $\Pi$ and $\Pi_{\mathrm{RPA}}$ is more dramatic because multiple excitations and decay processes are not accounted for at the RPA level. In the $\{ T, \eta \} \to 0$ limit there exist broad regions where $\mathrm{Im} \Pi_{\mathrm{RPA}}$ is identically zero, e.g. above the plasmon dispersion curve for $Q<Q_m$. In contrast, $\mathrm{Im}\Pi$ is non-zero in this spectral region, and in the $\Omega \to \infty$ limit $\mathrm{Im}\Pi \propto \Omega^{-3}$.

From the above asymptotic behavior and Eq. (5) of the main text it follows that for the exchange-correlation kernel $\mathrm{Re}K_{\mathrm{xc}}(\Omega \to \infty) \to B(Q,T)$ and $\mathrm{Im}K_{\mathrm{xc}}(\Omega \to \infty) \to 0$.